\documentclass[10pt,twocolumn]{article}

\usepackage{amsmath}
\usepackage{multirow}
\usepackage{graphicx}
\usepackage{subcaption}
\usepackage[margin=2cm]{geometry}
\usepackage{etoolbox}
\usepackage{adjustbox}
\usepackage{array}
\newcolumntype{C}[1]{>{\centering}m{#1}}
\usepackage{authblk}
\usepackage{upgreek}
\usepackage{xfrac}
\usepackage{booktabs}
\usepackage{color}
\usepackage[normalem]{ulem}
\usepackage[round,sort&compress,numbers]{natbib}

\title{Multifaceted design optimisation for superomniphobic surfaces}
\author[1]{Panter J. R.}
\author[2]{Gizaw, Y.}
\author[1]{Kusumaatmaja H.}
\affil[1]{Department of Physics, Durham University, South Road, Durham, DH1 3LE}
\affil[2]{Procter and Gamble Co., Winton Hill Business Center, 6210 Center Hill Avenue, Cincinnati, OH, United States}

\date{}
\begin{document}

\twocolumn[
\begin{@twocolumnfalse}
\maketitle

\begin{abstract}

Superomniphobic textures are at the frontier of surface design for vast arrays of applications. Despite recent significant advances in fabrication methods for reentrant and doubly reentrant microstructures, design optimisation remains a major challenge. We overcome this in two stages. Firstly, we develop readily-generalisable computational methods to systematically survey three key wetting properties: contact angle hysteresis, critical pressure, and minimum energy wetting barrier. For each, we uncover multiple competing mechanisms, leading to the development of new quantitative models, and correction of inaccurate assumptions in prevailing models. Secondly, we combine these analyses simultaneously, demonstrating the power of this strategy by optimizing structures that are well-suited to overcome challenges faced by two emerging applications: membrane distillation and digital microfluidics. As the wetting properties are antagonistically coupled, this multifaceted approach is essential for optimal design. When large surveys are impractical, we show that genetic algorithms enable efficient optimisation, offering speedups of up to 10,000$\times$.

\end{abstract}
\end{@twocolumnfalse}
]

\section{\textbf{Introduction}}

Superomniphobic surfaces show physical micro- or nano-texturing which enable even low surface tension liquids to remain suspended atop a vapour-filled surface structure. This vapour-suspended state is prized for its liquid-shedding abilities, enabling high droplet mobility and low viscous drag \cite{Yong2017}. These surfaces have significant potential to be transformative across a broad array of applications. These range from tackling current global-scale crises, via sustainable technologies for water purification \cite{Gonzalez2017,Ali2018} and anti-microbial surfaces in biomedicine \cite{Elbourne2017,Su2018}, through everyday applications such as anti-fingerprint coatings \cite{Belhadjamor2018} and packaging designed to reduce food waste \cite{Li2018}, to digital microfluidics as a versatile biological and chemical technology \cite{Freire2016}.

Two promising textures aimed to enable these technologies are the reentrant (`\textsf{T}-shaped') and doubly reentrant (`\texttt{T}-shaped') geometries. Naturally-occurring examples of these structures have been shown to imbue the cuticle of the springtail arthropod (Collembola) with superolephobic properties even for highly-wetting, pressurised liquids; whilst further exhibiting abrasion-resistance and anti-microbial abilities \cite{Nickerl2013, Hensel2013, Helbig2011, Hensel2015}. Recent breakthroughs in microfabrication techniques have also allowed these reentrant and doubly reentrant structures, as well as more complex textures, to be produced with $\rm{\mu}$m-scale resolution, including using 3D printing technology, fluidization of polymer micropillars, and lithographic methods \cite{Liu2018, Choi2017, Yun2018, Liu2014}.

Despite these highly versatile techniques, a large obstacle still exists to widespread development: it is not known how to design the surface structures to enable optimal performance in real-world applications. Successful superomniphobic designs must exhibit three key wetting properties: (1) a low contact angle hysteresis to maximise liquid mobility \cite{Tuteja2008}, (2) a high critical pressure, the maximum sustainable pressure at which the superoloephobic state is stable \cite{Wang2015}, (3) a high energetic barrier to failure, in which liquid infiltrates the surface texture and the high liquid mobility is lost \cite{Bormashenko2015}. 

The complex surface designs means that both computational and experimental studies are expensive and time-consuming to perform, and so are largely restricted in scope to considering only single wetting properties - never all three. This is highly problematic as the structural parameters of the design couple each wetting property, often antagonistically. For example, two effective ways to increase the critical pressure are to decrease the pillar-pillar separation, and decrease the system scale \cite{Tuteja2008}. However, decreasing the separation results in an increase in the contact angle hysteresis \cite{Butt2015}, whereas decreasing the system scale decreases the energetic barrier to the wetting transition due to the liquid-vapour interfacial area decreasing in squared proportion. Although this three-fold perspective has been introduced and advocated before, see for example \cite{Tuteja2008}, the true lowest-energy failure mechanisms have never been incorporated.

This work overcomes the optimisation challenges for superomniphobic wetting property design. We begin by developing computational strategies in order to systematically survey the affect of the structural parameters on the CAH, critical pressure, and minimum energy barrier individually (Section \ref{section:results}). These methods are highly general and can be applied to any conceivable surface design. This leads to the discovery of new mechanisms for the receding contact line and failure of the suspended state, as well as the development and validation of quantitative analytical models. We correct a number of inaccurate assumptions in prevailing models. In particular, we highlight a new capillary bridge model to replace the grossly inaccurate prediction for the optimal texture height in the critical pressure study. 

To illustrate the importance of multifaceted optimisation, we then consider two relevant example applications: water purification via membrane distillation, and droplet-based digital microfluidics (Section \ref{section:sim_opt}). Membrane distillation shows significant potential as a sustainable, low-energy water purification technology, capable of extracting potable drinking water from highly contaminated  water sources (see \cite{Gonzalez2017,Ali2018} for recent reviews). A significant challenge however is that oils readily foul the membranes, leading to a breakdown in device performance. Meanwhile, digital microfluidics is anticipated to enable re-usable, re-configurable, and material-efficient lab-on-chip devices \cite{Freire2016}. In this technology, the major challenge is that commonly-used but low surface tension solvents pin strongly to the surface, leading to drop immobilisation and device failure \cite{Ding2012}. We find here that the doubly reentrant surface geometry is ideally situated to meet these challenges, as it is robust to pressure even for highly-wetting or surfactant-contaminated liquids. 

In such complex surface design featuring many antagonistically-coupled wetting properties, we recognise that it is not always desirable to perform large-scale wetting property surveys. Thus, in Section \ref{section:sim_opt}, we develop a genetic algorithm to perform the simultaneous optimisation with high efficiency, offering a speedup of up to 10,000$\times$. This versatile approach is highly complimentary to recent innovations in complex surface microfabrication techniques \cite{Choi2017,Liu2014,Liu2018,Yun2018}, such that together, we anticipate a transformative approach to surface design.

\section{\textbf{Results and Discussions}} \label{section:results}
\subsection{\textbf{Contact angle hysteresis}}

\begin{figure}[!ht]
\includegraphics[width=0.45\textwidth]{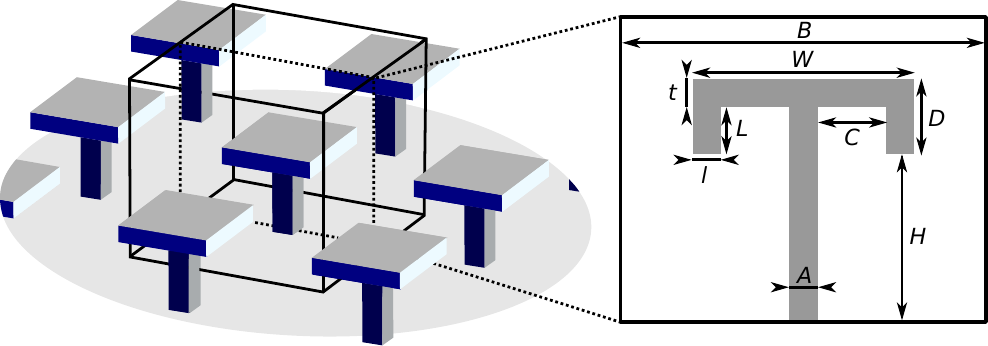}
\caption{Illustration of the 3D simulation repeat unit (left), with 2D cross section showing labelled structural parameters (right).}
\label{fig:1structure}
\end{figure}

We begin by simulating the liquid-vapour interface advancing and receding along a single row of surface structures (setup detailed in SI), in order to obtain the macroscopic advancing and receding contact angles $\theta_{a}$ and $\theta_{r}$ respectively, and the contact angle hysteresis (CAH = $\theta_{a} - \theta_{r}$). These simulated structures are shown in Fig. \ref{fig:1structure}, in which structures of variable dimensions are arranged in a square array. Throughout, all dimensions shown in Fig. \ref{fig:1structure} are reported relative to the system size $B$ and indicated with a subscript 'r'. For example the reduced cap width is $W_{\rm{r}} = W/B$. Unless otherwise stated, $B$ = 60 lattice spacings, with the cap thickness $t_{\rm{r}}$ and lip width $l_{\rm{r}}$ remaining fixed at 0.05. For the reentrant geometries, the lip depth $L_{\rm{r}} = 0$. Throughout, the microscopic contact angle $\theta_{\circ} = 60^{\circ}$ is used as a representative contact angle for organic solvents wetting fluorinated surfaces (see for example \cite{Lee2008}). We also investigate the non-wetting case, presented in SI.

\begin{figure*}[!ht]
\centering
\includegraphics[width=\textwidth]{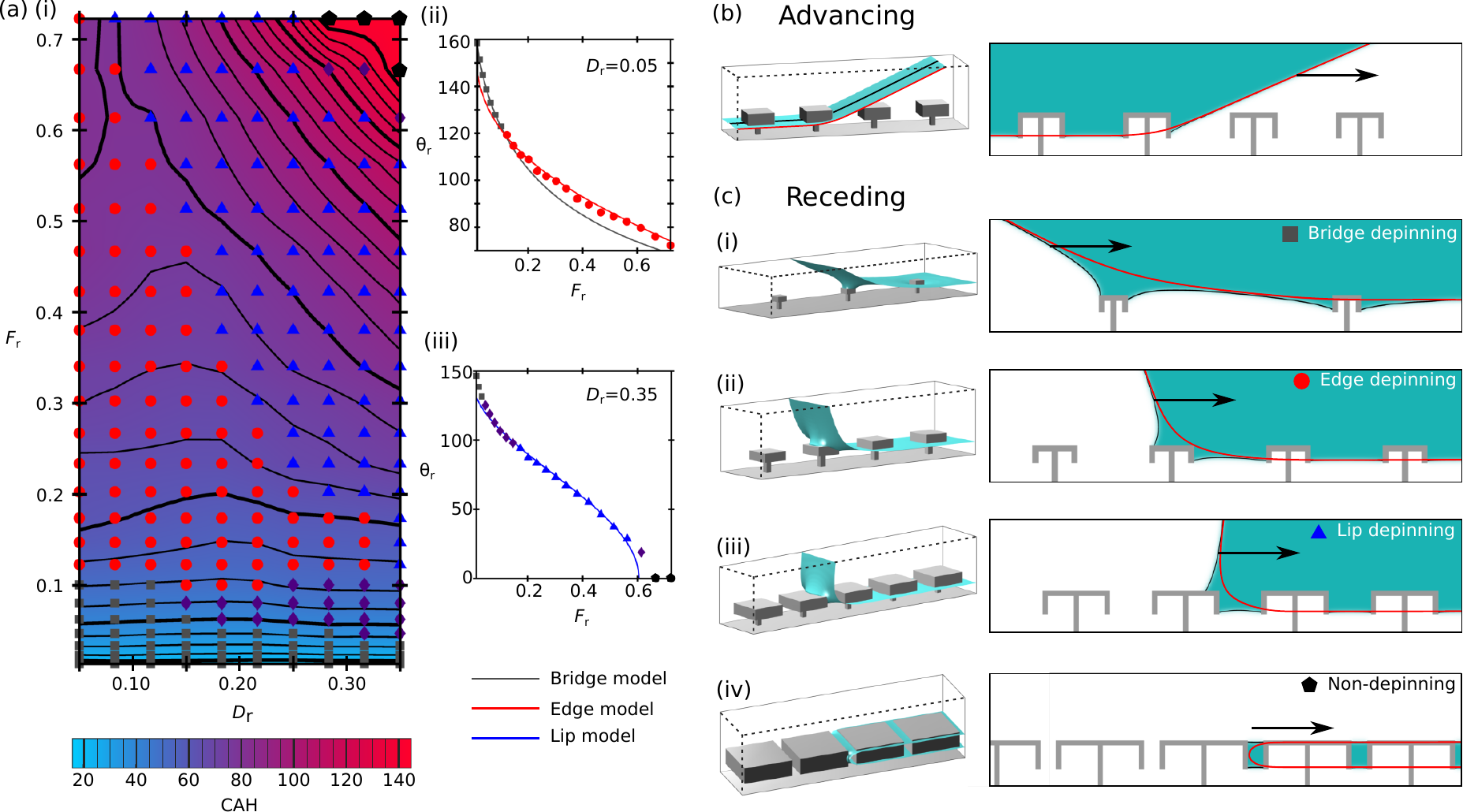}
\caption{Quantification and mechanisms leading to the contact angle hysteresis for reentrant and doubly reentrant geometries at zero applied pressure. (a)(i) CAH dependence on both the area fraction $F_{\rm{r}}$ and total cap height $D_{\rm{r}}$. Symbols indicate the depinning mechanism upon receding, with purple triangles indicating a hybrid mechanism. (ii), (iii) comparison of the bridge-, edge-, and lip-depinning receding models (solid lines, colour-coded) against the simulated $\theta_{\rm{r}}$ (data points); examples shown with varying $F_{\rm{r}}$ at fixed $D_{\rm{r}}$ = 0.05 and 0.35. The $\pm 1^{\circ}$ error bars in the simulation data are too small to be seen. (b) 3D visualisation of the advancing liquid-vapour interface (shown in blue), the advancing direction is indicated by a black arrow. Black and red lines indicate the centre and edge 2D cross sections which are also presented (right). (c)(i-iv) Visualisations of the major four receding mechanisms. The receding direction is indicated by black arrows.}
\label{fig:2CAH}
\end{figure*}

We find that the CAH depends only on the area fraction $F_{\rm{r}}$ of the cap ($F_{\rm{r}}=(W/B)^2$), and the total cap height $D_{\rm{r}}$, shown in Fig. \ref{fig:2CAH}(a)(i). Separate advancing and receding plots are shown in SI, alongside available comparison with previous experimental measurements. As the liquid-vapour interface never impinges under the cap, the hysteresis is identical for both reentrant and doubly reentrant geometries.

Across the simulated parameter range, the same advancing mechanism is observed, illustrated in Fig. \ref{fig:2CAH}(b). In this, the advancing occurs when the angle of the approximatively planar liquid-vapour interface results in the liquid contacting the cap of the neighbouring structure.

In contrast, the receding mechanism exhibits significant variation across the parameter range, and is therefore primarily responsible for the large variation in CAH observed in Fig. \ref{fig:2CAH}(a)(i). Overall, four dominant receding mechanisms are observed: bridge, edge, and lip depinning, and a fourth non-depinning mechanism. Characteristic examples of these are shown in Fig. \ref{fig:2CAH}(c)(i-iv) respectively. The operative regions of each mechanism are labelled in Fig. \ref{fig:2CAH}(a)(i), indicated by grey squares, red circles,  blue triangles and black pentagons respectively. The hybrid depinning mechanisms (purple diamonds), indicates the regions in which the dominant mechanisms smoothly interpolate. We now describe and model each of these receding mechanisms in turn.

For the lowest area fractions, at the point of receding, the three-phase contact line is pinned to the outermost pillar around the top perimeter of the cap, shown in Fig. \ref{fig:2CAH}(c)(i). A capillary bridge is strained between the cap and bulk liquid, such that receding occurs at the point of bridge depinning. This has been observed experimentally, \cite{Paxson2013,Butt2015,Butt2017}, and we are now able to  quantitatively test the receding model proposed by Butt \textit{et al.} \cite{Butt2015,Butt2017}. In this model, the receding liquid-vapour interface strains the capillary bridge parallel to the receding direction, thus tilting the bridge from the normal (detailed in the SI). The two suggested consequences of this are that the direction of the pinning force is tilted from the normal by $\pi/2 - \theta_{r}/2$, and that this force depends on the average contact angle $\bar{\theta} = \theta_{\rm{\circ}} + \pi/2 - \theta_{r}/2$. However, to yield an accurate model for use with wetting liquids as in this work, we stress that the appropriate average contact angle to use is $\bar{\theta} = \theta_{\rm{\circ}}^{\rm{max}} + \pi/2 - \theta_{r}/2$, where $\theta_{\rm{\circ}}^{\rm{max}} = max \left( \theta_{\rm{\circ}},\pi/2  \right)$. This is because for wetting liquids, the maximum pinning force is achieved when the contact angle reaches $90^\circ$. With this modification, at the point of receding this microscopic pinning force balances the macroscopic force required to move the contact line, yielding
\begin{equation}
\tan\left(\frac{\theta_r}{2}\right) = \frac{2}{4 W_{\rm{r}}\alpha}, \label{eqn:rec_bridge}
\end{equation}
In the simplest model, it is assumed that at the point of depinning, the three-phase contact line closely follows the square cap perimeter of length $4 W_{\rm{r}}$. However, to reflect the actual contact line morphology, the shape parameter $\alpha$ is introduced.  $\alpha$ is equal to 1 if the contact line is perfectly square, and $\uppi/4$ for a circle. As the shape parameter cannot be predicted \textit{a priori}, it is treated as a fitting parameter. An example of this is shown in Fig. \ref{fig:2CAH}(a)(ii) (grey curve and square points) for $D_{\rm{r}}$ = 0.05. Here, $\alpha$ = 0.861, reflecting the contact line deviating from perfectly square by depinning at the cap corners. This yields an average agreement between the simulation and model of $0.4^{\circ}$ (average agreement for all $D_{\rm{r}}$ tested is $0.6^{\circ}$, maximum  $2^{\circ}$).

At large $D_{\rm{r}}$ and $F_{\rm{r}}$, a new capillary bridge depinning mechanism is observed in which the bridge is strained between the cap edge, and bulk liquid phase. This lip-depinning receding mechanism, shown in Fig. \ref{fig:2CAH}(c)(iii), results in a substantial decrease in $\theta_{\rm{r}}$, demonstrated in Fig \ref{fig:2CAH}(a)(iii) (blue curve and triangular points). The model we introduce is to approximate the receding interface as a capillary bridge pinned to the side of the cap, stretched \textit{parallel} to the receding direction (detailed in SI). As with the bridge-depinning model however, we must account for the average contact angle around the contact line, yielding
\begin{equation}
\cos\left(\frac{\theta_r}{2}\right) = \frac{ 2 \left( W_{\rm{r}} + D_{\rm{r}}  \right) \alpha}{2}. \label{eqn:rec_lip}
\end{equation}
The accuracy of this model is demonstrated in \ref{fig:2CAH}a(iii) at $D_{\rm{r}}$ = 0.35, for which $\alpha$ = 0.887. This yields an average agreement between the simulation and model of $1^{\circ}$, (average agreement for all $D_{\rm{r}}$ tested is $2^{\circ}$, maximum  $6^{\circ}$).

This lip-depinning model also predicts the existence of systems in which a receding contact angle no longer exists. In these extreme cases of the lip-depinning mechanism, because depinning is not able to occur, a droplet caused to move across the surface would leave a trail of suspended liquid trapped between the caps. We observe this predicted non-depinning mechanism in simulations, indicated by black pentagons in Figs. \ref{fig:2CAH}(a)(i,iii). This non-receding case is illustrated in Fig. \ref{fig:2CAH}(c)(iv).  

At intermediate area fractions and lip depths, the depinning mechanism is no longer capillary bridge-mediated. Instead, the edge-pinned receding mechanism is observed, shown in Fig. \ref{fig:2CAH}(c)(ii). Here, the interface maintains approximately the same morphology as it depins laterally from the edge of the cap. Thus, we are able to analyse the energetic change of sliding the interface laterally by a small distance, in order to obtain the angle at which receding becomes energetically favourable - the receding angle. This is derived in SI. Taking into account the liquid receding from the cap top and edges, and top surface of the microchannel, 
\begin{equation}
\cos\theta_r =  (W_{\rm{r}} + 2D_{\rm{r}})\cos\theta_{\circ} + W_{\rm{r}} -1. \label{eqn:rec_edge}
\end{equation}
This represents a generalisation of previous edge-depinning models \cite{Choi2009,Mognetti2010}, in which by taking account of the liquid receding from the cap sides, we are now able to describe the edge-depinning mechanism accurately for wetting liquids. This is demonstrated in Fig. \ref{fig:2CAH}(a)(ii) (red curve and circular points). Without any fitting parameters, the average agreement between the model and simulation results is $2^{\circ}$. (average agreement for all $D_{\rm{r}}$ tested is $3^{\circ}$, maximum  $7^{\circ}$).

\subsection{\textbf{Critical pressure}}

\begin{figure*}[!ht]
\centering
\includegraphics[width=\textwidth]{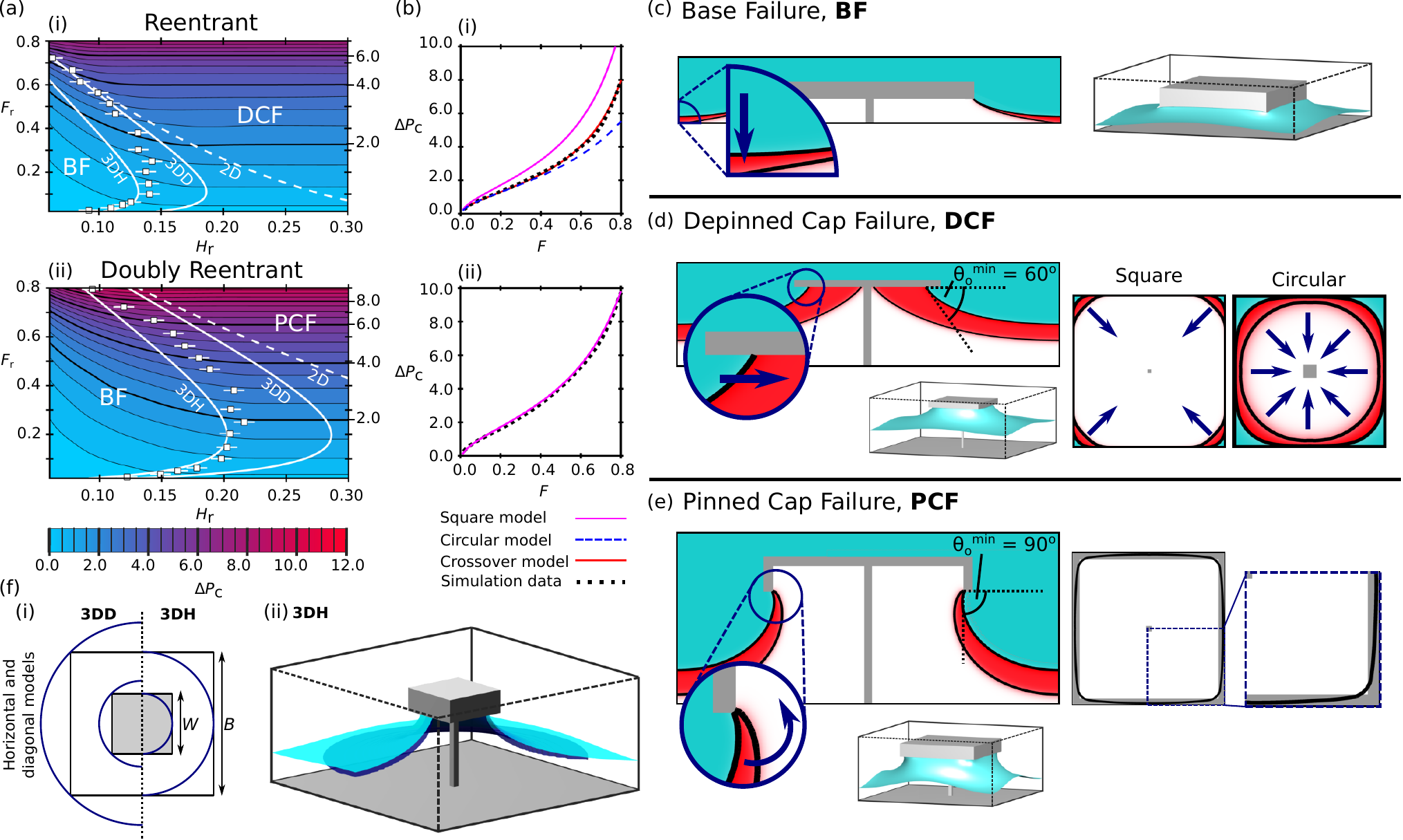}
\caption{Critical pressure analysis for reentrant and doubly reentrant geometries. (a) Contour plots of $\Delta P_{\rm{c}}$ variation with $F_{\rm{r}}$ and $H_{\rm{r}}$ for reentrant (i) and doubly reentrant (ii) geometries. Data points mark the critical height at which the failure mechanism switches from Base Failure (BF) to cap failure (CF), error bars indicate the uncertainty in this height due to the diffuse interface width. Solid and dashed white lines show the critical height based on the capillary model and 2D model respectively. (b) Model fits to $\Delta P_{\rm{c}}$ of the cap failure mechanisms at $H_{\rm{r}}$ = 0.25 for reentrant (i) and doubly reentrant (ii) geometries. (c-e) The three failure mechanisms shown in 3D, with associated diagonal cross sections. Critical pressure liquid morphologies are shown in blue, the vapour phase shown in white, and the interface indicated with a black solid line. Red regions show how the unstable meniscus evolves upon increasing $\Delta P$ above $\Delta P_{\rm{c}}$. Panels (d,e) also show under-cap views, highlighting the shapes of the contact lines at the critical pressure. (f) Details of the horizontal and diagonal capillary bridge models used, showing the inner and outer circumferences (blue) against the system configuration. The 3D illustration compares the simulated liquid-vapour interface (light blue) to the horizontal capillary model (dark blue).}
\label{fig:3PCRIT}
\end{figure*}

Unlike with the contact angle hysteresis, the critical pressure is sensitive to whether the surface geometry is reentrant or doubly reentrant. Throughout, the critical pressure shown $\Delta P_{\rm{c}}$ is referenced with respect to the pressure $\upgamma_{\rm{lv}}/B$. Here we find that $\Delta P_{\rm{c}}$ is only dependent upon the area fraction $F_{\rm{r}}$ and the pillar height $H_{\rm{r}}$. Although $\Delta P_{\rm{c}}$ is affected by the presence of a doubly reentrant lip, $\Delta P_{\rm{c}}$ does not depend on the precise lip depth $L_{\rm{r}}$. The critical pressure dependencies on $F_{\rm{r}}$ and $H_{\rm{r}}$ are shown for the reentrant and doubly reentrant structures in Figs. \ref{fig:3PCRIT}(a)(i,ii) respectively. For both structural types, these dependencies change across the parameter space, due to the presence of two different pressure-induced failure mechanisms: Base Failure and Cap Failure.  

For a given area fraction $F_{\rm{r}}$, at low pillar heights the Base Failure mechanism is operative, illustrated in Fig. \ref{fig:3PCRIT}(c). In this, the suspended state fails because the sagging liquid-vapour interface touches the base of the system whilst the three-phase contact line remains pinned to the bottom of the cap. By increasing the pillar height, $\Delta P_{\rm{c}}$ is increased.

However, at $H_{\rm{r}}$ above the critical height, $H_{\rm{c}}$, the Cap failure mechanism becomes operative. Here, at the critical pressure the system can no longer simultaneously support the uniform mean curvature of the liquid-vapour interface and the contact line morphology. For the reentrant geometry, this results in the contact line depinning and sliding inwards, shown in Fig. \ref{fig:3PCRIT}(d). For the doubly reentrant geometry with a thin lip width $l_{\rm{r}}$, this results in the liquid-vapour interface ballooning outwards while the contact line remains pinned, shown in fig. \ref{fig:3PCRIT}(e). In both of these cases, increasing the pillar height further now results in no change to $\Delta P_{\rm{c}}$.

Therefore at fixed $F_{\rm{r}}$, the maximum $\Delta P_{\rm{c}}$ occurs for $H_{\rm{r}} \geq H_{\rm{c}}$, in the Cap failure region. However, it is detrimental for design performance if the height is increased above the critical height, as this mechanically weakens the structure without increasing $\Delta P_{\rm{c}}$ \cite{Zhao2012,Dyett2014}. The optimum pillar height is therefore $H_{\rm{r}} = H_{\rm{c}}$, which defines the Base Failure - Cap Failure boundary. We therefore focus on discussing the critical pressure due to the Cap failure mechanisms, before analysing the critical height.

\subsubsection{\textbf{Depinned Cap Failure for Reentrant Geometries}}
In order to understand how $\Delta P_{\rm{c}}$ is influenced by the area fraction (or alternatively the cap width $W_{\rm{r}}$) in the depinned Cap Failure mechanism, we begin by examining the rudimentary model proposed by Tuteja \textit{et al.} \cite{Tuteja2007, Tuteja2008}. In this, 
\begin{equation}
\Delta P_{\rm{c}} = \frac{4 \alpha \sin{\theta_{\circ}}}{\frac{1}{W_{\rm{r}}} - \alpha{W_{\rm{r}}}}. \label{eqn:pcrit}
\end{equation}
For convenience, we incorporate the shape parameter $\alpha$, the same parameter as defined previously in the CAH section, in order to unify the critical pressure models on circular ($\alpha = \uppi/4$) and square ($\alpha = 1$) geometries. To rationalise this model, at $\Delta P_{\rm{c}}$ the pinning force of the contact line balances the force due to the pressure over the area between the pillars \cite{Tuteja2007, Tuteja2008}. Two key assumptions are made. Firstly, the contact line is supposed to follow the cap edge, whilst secondly, the contact angle around the contact line is presumed to be uniform and equal to $\theta_{\circ}$. We test this model in Fig. \ref{fig:3PCRIT}(b)(i). The square-cap model is observed to fit the simulation results very poorly, overestimating the critical pressure by between 26\% to 95\% in the tested range $0.016 \leq F_{\rm{r}} \leq 0.8$. If instead, a circular contact line model is employed, we find that this agrees with the simulation data up to moderate area fractions ($F_{\rm{r}} < 0.6$). Overall, by observing the contact line shape obtained through simulations, shown in Fig. \ref{fig:3PCRIT}(d), we find that the contact line varies in morphology, from circular at low $F_{\rm{r}}$, to approximately square at high $F_{\rm{r}}$.

We now develop a more sophisticated model, capable of accurately describing the critical pressure for reentrant and doubly reentrant geometries, at all contact angles. Three modifications are introduced to Eq. \eqref{eqn:pcrit}. Firstly, it cannot be assumed that the contact line follows the cap edge, leading to the introduction of $W_{\rm{r}}'$, the corrected reduced width: $W_{\rm{r}}' = W_{\rm{r}} - a$, where $a$ is a parameter which describes the difference between the actual width that the contact line assumes and the width of the cap. Since $a$ cannot be predicted  \textit{a priori} we treat $a$ as a second fitting parameter. For reentrant geometries, and doubly reentrant geometries with 
 $\theta_{\circ} > 90^{\circ}$, we anticipate $a \approx 0$, due to contact line pinning on the outer edge. For doubly reentrant geometries with  $\theta_{\circ} < 90^{\circ}$, we anticipate $a \approx 2L_{\rm{r}}$, due to contact line pinning on the inner edge. Secondly,  we propose that the shape parameter $\alpha$ varies continuously as a function of $W_{\rm{r}}'$ between the circular and square limits, such that  $\alpha = \frac{\uppi}{4} + (1- \frac{\uppi}{4})(W_{\rm{r}}')^{x}$. The exponent $x$ describes the strength of this crossover, and is a second fitting parameter. Thirdly, if the contact angle on the hydrophilic reentrant geometry is increased, the pinning force of the contact line is maximised at $\theta_{\circ}=90^{\circ}$. For $\theta_{\circ} > 90^{\circ}$, the pinning force remains at this maximal value (shown in SI). Thus, to generally describe the critical pressure on all reentrant geometries, we replace $\theta_{\circ}$ in Eq. \eqref{eqn:pcrit} by the corrected contact angle $\theta_{\circ}^{\rm{min}} = \min[\theta_{\circ}, 90^{\circ}]$. 

This crossover model is shown in Fig. \ref{fig:3PCRIT}(b)(i) to be in excellent agreement with the simulation data, yielding $ a = 0.023$ and $ x = 6.7$. As anticipated, $a$ is small relative to the cap width, and is of the order of the diffuse interface width ($\epsilon/B = 0.01$). The large exponent $x$ reflects the simultaneous change of both the perimeter and area of the contact line as the system crosses from a circular to square configuration.

\subsubsection{\textbf{Pinned Cap Failure for Doubly Reentrant Geometries}}
Next, we consider the critical pressures of the doubly reeentrant structure, shown in Fig. \ref{fig:3PCRIT}(b)(ii). In the pinned Cap Failure region, the liquid-vapour interface is pinned to the inner cap edge, shown in Fig. \ref{fig:3PCRIT}(e). The doubly reentrant lip enforces an approximately square contact line across the entire range of $F_{\rm{r}}$ tested, such that excellent agreement between the simulations and the critical pressure model in Eq. \eqref{eqn:pcrit} is achieved at $x=0$, $\alpha=1$ and $a = 0.080$. This model is also successfully employed for $\theta_{\circ} = 110^{\circ}$ in SI. As anticipated, within the uncertainty introduced by the diffuse interface $a \approx 2L_{\rm{r}}$ (where $2L_{\rm{r}}=0.1$). However, the data in Fig. \ref{fig:3PCRIT}(b)(ii) are best described by replacing $\theta_{\circ}$ in Eq. \eqref{eqn:pcrit} with $90^{\circ}$. In 2D, and for axisymmetric doubly reentrant wells, it is facile to show that the critical pressure occurs when the contact angle reaches $90^{\circ}$ for thin lip widths $l_{\rm{r}}$ \cite{Hensel2013}. We conclude here that this remains true, even for 3D, non-axisymmetric caps. This therefore verifies the proposition that the doubly reentrant lip maximises the critical pressure for any surface wettability \cite{Liu2014}.

\subsubsection{\textbf{Critical heights}}
At a given $F_{\rm{r}}$, the critical heights are obtained both analytically and through simulations by observing the maximum depth which the liquid-vapour interface sags under the pillar in the Cap failure regions. If the pillar height $H_{\rm{r}}$ is equal to this sagging depth, the failure mechanism is simultaneously Base Failure and Cap failure, so defining the failure mode boundary and the critical height $H_{\rm{c}}$. The salient observation based on the simulated critical heights (white data points in Figs. \ref{fig:3PCRIT}(a)(i,ii)) is that regardless of cap area fraction, the optimal pillar height is surprisingly short, and should never exceed $\approx0.2B$. We now rationalise this observation.

In many critical pressure models, a 2D circular-arc model is employed to estimate the sagging height of the liquid-vapour interface,
\begin{equation}
H_{\rm{c}} = \frac{S_{\rm{r}}}{2} \frac{1-\cos{\theta_{\circ}^{\rm{min}}}}{\sin{\theta_{\circ}^{\rm{min}}}}, \label{eqn:hcrit}
\end{equation}
where the separation $S_{\rm{r}} = (1-W_{\rm{r}})$ between horizontally adjacent pillars (see for example \cite{Tuteja2007, Tuteja2008}), or $S_{\rm{r}} = \sqrt{2}(1-W_{\rm{r}})$ for diagonally separated pillars (see for example \cite{Butt2013}). The latter model is shown in Figs. \ref{fig:3PCRIT}(a)(i,ii) (dashed white lines), and exemplifies the conclusion that, except at very high $F_{\rm{r}}$, a 2D estimation grossly overestimates the critical height. All currently manufactured, low-$F_{\rm{r}}$ structures relying on these 2D models are therefore significantly taller than necessary, which can be seen in refs. \cite{Wang2015} and \cite{Grigoryev2012} for example.

The actual, non-monotonic critical height variation with $F_{\rm{r}}$, shown in Figs. \ref{fig:3PCRIT}(a)(i,ii), can be rationalised by considering that in 3D, two principal radii of curvature characterise the liquid-vapour interface at each point. At low $F_{\rm{r}}$, the small contact line radius enforces a small, negative principal radius if curvature $R_{\rm{1}}$ on the liquid-vapour interface. Since the critical pressure is positive, and proportional to $ 1/R_{\rm{1}} + 1/R_{\rm{2}}$, the second principal radius of curvature $R_{\rm{2}}$ must be smaller in magnitude than $R_{\rm{1}}$, and positive, resulting in a significantly reduced sag height compared to the 2D case.  At large $F_{\rm{r}}$ however, observable for the reentrant geometry in Fig. \ref{fig:3PCRIT}(a)(i), the 2D model is recovered as the principal radius of curvature approximated by the circular arc ($R_1$) is significantly smaller than the second principal radius of curvature of the interface ($R_2$). In this case the interface shape becomes well-approximated by the single radius of curvature $R_1$.

We therefore recognise that the liquid-vapour interface is able to be modelled as a capillary bridge, for which we define the inner radius to contact the cap at an angle $\theta_{\circ}^{\rm{min}}$, and the outer radius to contact the simulation boundary with an angle equal to zero. Because the capillary bridge is axisymmetric, whereas the simulated system is square, there are two limiting cases of where the inner and outer radii contact the structure and simulation boundary respectively, shown in Fig. \ref{fig:3PCRIT}(f)(i). Either the inner and outer radii contact the central edges of the structure and simulation boundaries (the horizontal model, representing the minimum possible radii), or contact is made at the corners (the diagonal model, representing the maximum possible radii). Both potential capillary models are shown in Figs. \ref{fig:3PCRIT}(a)(i,ii) (solid white lines). We reserve the detailed derivation to the SI. 

In all cases it is observed that the simulated critical heights are bounded by the horizontal and diagonal capillary bridge models. Furthermore, at low $F_{\rm{r}}$, the horizontal capillary model accurately predicts the critical height. For structures where the interface is pinned to the outer cap edge, namely reentrant geometries and doubly reentrant geometries with $\theta_{\circ} > 90^{\circ}$ (shown in SI), the diagonal model is shown to closely predict the critical height at high $F_{\rm{r}}$. An illustrated comparison of the horizontal and diagonal models is shown in Fig. \ref{fig:3PCRIT}(f)(ii). Through this, we are able to successfully capture the critical height suppression at low $F_{\rm{r}}$, and maximum $H_{\rm{c}}$ at intermediate $F_{\rm{r}}$. We further validate the capillary bridge model in SI by showing that this model is able to accurately reproduce an experimental interfacial profile from \cite{Papadopoulos2013}, whereas a circular arc model significantly overestimates the interfacial sagging height.

\subsection{\textbf{Minimum energy transition mechanisms}}

\subsubsection{Transition states and pathways}

\begin{figure*}[!ht]
\includegraphics[width=\textwidth]{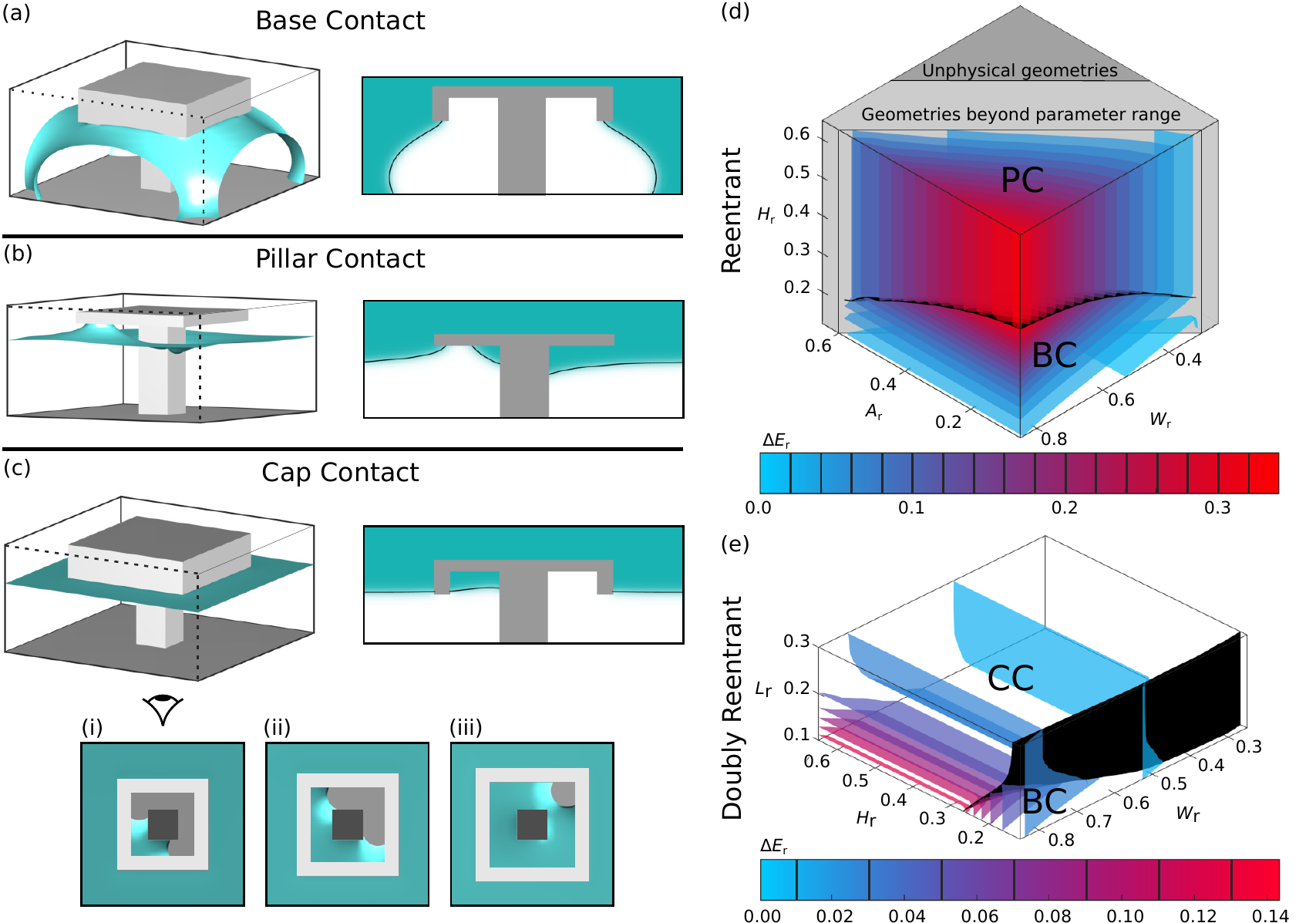}
\caption{(a-c) 3D visualisations of the transition states of each transition pathway (liquid-vapour interfaces shown in blue) with associated diagonal cross sections (liquid-vapour interface outlined in black). In panel (c), under-cap views show the three Cap Contact transition state morphologies. (d) 3D contour plots showing the energy barrier $\Delta E_{\rm{r}}$ of the lowest energy transition mechanism for the reentrant geometries. Each surface is a surface of constant $\Delta E_{\rm{r}}$. The dividing surface between different transition mechanisms is shown in black. Impossible geometries with pillar widths $A_{\rm{r}}$ wider than the cap width $W_{\rm{r}}$ are shaded in dark grey. Geometries approaching this limit and requiring infeasibly large computational domains are shaded in light grey. (e) 3D contour plots showing the energy barrier $\Delta E_{\rm{r}}$ of the lowest energy transition mechanism for the doubly reentrant geometries.}
\label{fig:4TS}
\end{figure*}

In order to design surfaces which maintain a suspended state in challenging environments, it is not sufficient to only understand how susceptible a surface design is to pressure. Failure can be initiated through a broad range of additional perturbations, such as: flow\cite{Zhang2017}, vibration\cite{Papadopoulos2016}, evaporation\cite{Susarrey-Arce2012}, condensation \cite{Dorrer2007}, droplet impact \cite{Zhang2015}, changes to electric\cite{Chen2011} or magnetic\cite{Grigoryev2012} fields, or thermal fluctuations at the nanoscale\cite{Amabili2016}. In a real application, several perturbations will be present simultaneously meaning that failure is unlikely to be initiated by only a single perturbation, but instead via their combination. In fabricating a texture resistant to failure, it is therefore vital to understand this combined failure in the worst-case scenario - the minimum-energy pathway by which the suspended state collapses. This is a steepest-descent pathway between two metastable states, in which the maximum energy along the path occurs at a saddle point (the transition state). The minimum energy barrier is the difference in energies between the suspended and transition states. This places a lower bound on the collapse energy barrier. If this barrier cannot be overcome by the perturbations applied to a candidate surface design, the suspended state can be guaranteed to remain stable. 

Through utilising the Doubly Nudged Elastic Band algorithm \cite{Trygubenko2004} three transition pathways are found: Base Contact, Pillar Contact, and Cap Contact, visualised in movies S1, S2, and S3 respectively. From these, the transition state morphologies are surveyed using the gradient-squared method, each of which are shown in Figs. \ref{fig:4TS}(a-c). In the large scale structural surveys, as we only wish to obtain the transition states and minimum energy barriers rather than the full paths, the gradient squared method is significantly more efficient that the full pathway algorithm, as only a single minimisation is required, as opposed to evolving a string of multiple systems across the landscape. These transition search algorithms are detailed further in SI. We begin by discussing the qualitative characteristics of each before quantifying the suspended-to-collapsed minimum energy barrier. All transition state searches are carried out at zero applied pressure. 

The Base Contact mechanism (BC), shown in Fig. \ref{fig:4TS}(a), is observed for both reentrant and doubly reentrant geometries. The mechanism proceeds via the liquid-vapour interface sagging towards the system base whilst pinned to the cap lip. The transition state is observed to occur after the liquid meniscus has contacted the base of the system. This mechanism is highly prevalent across a broad range of structural and chemical properties, such on non-wetting geometries (see SI) as well as pillars and nails \cite{Panter2017, Savoy2012, Zhang2014, Pashos2016, Amabili2016}.

The Pillar Contact mechanism (PC), illustrated in Figs. \ref{fig:4TS}(b), occurs only for the reentrant geometry. The mechanisms proceeds via the liquid-vapour interface impinging under the cap, with the transition state occurring before the interface detaches from the solid surface. Following this, the interface slides down the pillar to contact the base and complete the transition. 

Finally, the Cap Contact mechanism (CC) is observed only on the doubly reentrant geometries, shown in Fig. \ref{fig:4TS}(c). Here, the expected Pillar Contact mechanism is unstable with respect to condensation of liquid inside the cap structure. In this mechanism, the transition begins with the condensation of liquid in one corner of the cap underside, which subsequently grows to fill the cap. The precise location of the transition state can take one of three morphologies, shown in Fig. \ref{fig:4TS}(c)(i-iii). Which variant occurs is discussed in the SI. The transition then continues by filling the cap entirely, such that a new free energy minimum is obtained, a suspended state with a liquid-filled cap. An additional energy barrier is required to complete the wetting mechanism: the two separate liquid-vapour interfaces must coalesce by crossing the cap lip, before the remainder of the mechanism proceeds exactly as with Pillar Contact. However, this coalescence barrier is small relative to the condensation barrier presented, and decreases further as the lip width is reduced. This mechanism is particularly important to understand as for many applications requiring the doubly reentrant geometry, the liquids are volatile, or else the suspended state needs to be maintained over long time-scales. Two such applications are discussed further in the Simultaneous Optimisation section. 

Of further note is that the transitions presented here are all MEPs regardless of liquid volatility. The pressure treatment in the free energy functional used in these simulations effectively contacts every point of the system with an external fluid reservoir at constant pressure. Thus, fluids may be exchanged anywhere within the system. For non-volatile liquids, although the Cap Contact condensation mechanism is a minimum energy pathway on the doubly reentrant geometry, CC may not be realised on an experimental timescale. The transition will therefore occur via a non-condensing route, the minimum energy path of which is Base Contact.

\subsubsection{\textbf{Minimum energy barriers}}
\begin{table}
\scriptsize
\begin{adjustbox}{width=0.5\textwidth}
	\begin{tabular}{C{0.5cm}| C{1.1cm}| C{1.1cm}|| C{1.1cm}| c}

	& \multicolumn{2}{c||}{\textbf{Reentrant}} &  \multicolumn{2}{c}{\textbf{Doubly Reentrant}} \\
	& BC & PC & BC & CC \\
 \hline
	$H_{\rm{r}}$ & \normalsize{$\bullet$} & & \normalsize{$\bullet$} & \\
	$W_{\rm{r}}$ & \normalsize{$\bullet$} & \normalsize{$\bullet$}& \normalsize{$\bullet$} & \normalsize{$\bullet$} \\
	$A_{\rm{r}}$ &  & \normalsize{$\bullet$} &  & \normalsize{$\bullet$} \\
	$L_{\rm{r}}$ & & & & \normalsize{$\bullet$}  \\
	\bottomrule
	\end{tabular}
	\end{adjustbox}
\caption{The geometrical parameters which affect $\Delta E_{\rm{r}}$ for each transition mechanism, indicated with filled circles.}
\label{table:TStype}
\end{table}

Overall, we find that each structure has at most two potential transition pathways. Throughout this work, the energy barrier $\Delta E_{\rm{r}}$ is expressed relative to the reference energy $\upgamma_{\rm{lv}}B^2$. The barrier of each pathway is affected differently by the structural parameters, which we summarise in Table \ref{table:TStype}. In the style of traditional phase diagrams, we present the lowest energy-barrier mechanism at each parameter value tested in Figs. \ref{fig:4TS}(d,e) and so are able to predict the dominant collapse mechanism. 

Beginning with the reentrant geometry, Base and Pillar Contact compete for the lowest energy collapse mechanism, shown in Fig. \ref{fig:4TS}(d). As shown in Table \ref{table:TStype}, for all Base Contact transition states, the interface morphology depends on the pillar height $H_{\rm{r}}$ and cap width $W_{\rm{r}}$. As both of these structural parameters increases, the liquid-vapour interface increases in area leading to an increase in $\Delta E_{\rm{r}}$. However,  $\Delta E_{\rm{r}}$ cannot be increased indefinitely by increasing $H_{\rm{r}}$, as at a critical pillar height, the height-independent Pillar Contact mechanism becomes the lowest energy transition pathway. As the Pillar Contact transition state is associated with the liquid wetting the reentrant cap underside, $\Delta E_{\rm{r}}$ is increased by expanding the liquid-vapour interfacial area required to do so. This requires $W_{\rm{r}}$ to be maximised, and the pillar width $A_{\rm{r}}$ to be minimised. In designing a reentrant structure exhibiting the maximum energy barrier, the mechanistic switch upon increasing $H_{\rm{r}}$ is a key point to highlight as, assuming taller pillars are mechanically weaker than shorter pillars \cite{Zhao2012,Dyett2014}, at a given $W_{\rm{r}}$ and $A_{\rm{r}}$ the optimal structure height is that on the BC-PC boundary, reminiscent of the critical pressure case. 

Fig. \ref{fig:4TS}(e) shows the lowest energy collapse mechanism for the doubly reentrant geometries,  which is dramatically different to the (singly) reentrant equivalent. Here, it is the Base Contact and Cap Contact mechanisms which compete for the lowest energy path. In Table \ref{table:TStype}, we find that the minimum energy barrier depends on four parameters: the Cap Contact energy barrier depends on $W_{\rm{r}}$, $A_{\rm{r}}$, and the lip depth $L_{\rm{r}}$,  whereas the Base Contact barrier depends on $W_{\rm{r}}$ and also $H_{\rm{r}}$. However, ubiquitously the lowest Cap Contact barriers are obtained by minimising $A_{\rm{r}}$, such that the barrier diagram shown in Fig. \ref{fig:4TS}(e) is at constant $A_{\rm{r}} = 0.05$.

Regarding the Base Contact mechanism, the only effect on changing from a reentrant to doubly reentrant geometry is to increase the range of $W_{\rm{r}}$ and $H_{\rm{r}}$ for which the Base Contact mechanism is operative. However, except for the smallest pillar heights, the Cap Contact mechanism has a significantly smaller $\Delta E_{\rm{r}}$ compared to the Base Contact barrier. This is principally caused by the condensing critical nucleus having a relatively small, energetically unfavourable liquid-vapour interfacial area, compared to the large, energetically favourable solid-liquid interfacial area. Therefore to maximise the Cap Contact barrier, the liquid-vapour interfacial area must be maximised, whilst minimising the solid-liquid contact area. This is effectively realised by maximising $W_{\rm{r}}$ and minimising $L_{\rm{r}}$ and $A_{\rm{r}}$. As the Cap Contact mechanism is independent of the height of the structure, in a similar manner to the reentrant structure the optimal pillar height is located on the boundary between the two failure mechanisms.

\section{\textbf{Simultaneous optimisation}}  \label{section:sim_opt}

\begin{figure*}[!t]
\centering
\includegraphics[width=1.0\textwidth]{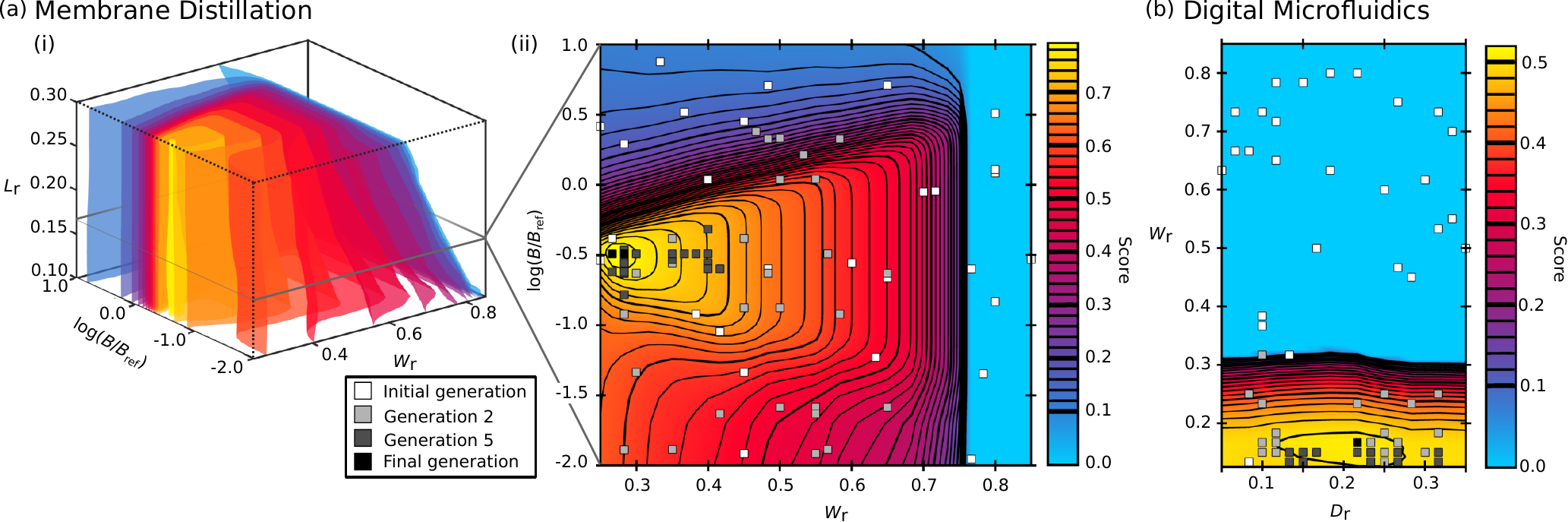}
\caption{(a)(i) 3D contour plot of the membrane distillation scoring function at fixed $H_{\rm{r}} = 0.3$, $A_{\rm{r}} = 0.05$,  $t_{\rm{r}} = 0.05$. Each surface is a surface of constant score. (ii) A 2D slice of the 3D contour plot at the optimal $L_{\rm{r}} = 0.17$. Square data points show the initial (white), 2nd (light grey), 5th (dark grey), and final (black) generations of the genetic algorithm, projected onto the 2D plane. (b) Scoring function for the digital microfluidics application, projected onto the $H_{\rm{r}} = 0.3$ plane at fixed $B = 100$ $\upmu$m, also showing the successive generations of the genetic algorithm population.}
\label{fig:SIM_OPT}
\end{figure*}

Overall, six structural parameters influence the three key wetting properties:  $A_{\rm{r}}$ (pillar width), $H_{\rm{r}}$ (pillar height), $ L_{\rm{r}}$ (lip depth), $t_{\rm{r}}$ (cap thickness), $W_{\rm{r}}$ (cap width), and the system scale $B/B_{\rm{ref}}$ (where $B_{\rm{ref}}$ = 1 $\upmu$m). Having studied how these parameters affect each individual wetting property, the parameters which antagonistically couple the wetting properties become apparent. Firstly, in order to reduce the CAH, $W_{\rm{r}}$ must be reduced; but this reduces $\Delta P_{\rm{c}}$ and $\Delta E_{\rm{r}}$. Secondly, in order to increase $\Delta P_{\rm{c}}$ the system scale must be reduced; but this reduces $\Delta E_{\rm{r}}$. In order to overcome this unfavourable coupling, we simultaneously optimise the surface structures, which is demonstrated for two example applications: membranes for water purification via membrane distillation, and droplet-based digital microfluidics. In order to perform this simultaneous optimisation, we begin by developing an application-specific scoring function which grades a candidate design against the desired wetting properties. We then optimise the scoring function using two methods. The first is to evaluate the scoring function over the entire parameter range tested, and from this find the optimum structure. In the second method, we demonstrate that for designs where it is not practical to perform comprehensive wetting property surveys, due to increased surface complexity for example, genetic algorithms can be used to efficiently perform the simultaneous optimisation.

Conventionally in membrane distillation, purification is achieved by passing a heated contaminated water source over a hydrophobic membrane, through which water vapour is able to pass to collect in a clean water reservoir. However oils readily foul the membrane leading to breakthrough of the contaminated liquid into the fresh water reservoir \cite{Gonzalez2017,Wang2017}. We overcome this using a doubly reentrant structure. To optimise the geometry, we construct a suitable scoring by recognising that the first priority is for the membrane to be pressure-resistant under typical operating conditions: water at $70^{\circ}$C ($\upgamma_{\rm{lv}} = 64.4$ mN$\cdot$m$^{-1}$) under pressures of approximately 100 kPa ($\Delta P_{\rm{c}}^{\rm{target}}$) \cite{Alkhudhiri2012}. In order to ensure the suspended state remains stable, the minimum energy barrier must be of the order of 100 k$_{\rm{B}}$T ($\Delta E_{\rm{r}}^{\rm{target}}$). This is of particular importance for the doubly reentrant geometry, in order to prevent failure via condensation within the texture \cite{Xue2016}. Finally, in order to reduce viscous drag across the membrane the CAH should be minimised, and we impose the condition that the CAH should not exceed 90$^{\circ}$ ($\rm{CAH}^{\rm{cutoff}}$). The critical pressure, energy barrier, and CAH conditions generate individual scoring functions $S_{\rm{P}}$, $S_{\rm{E}}$, and $S_{\rm{C}}$ respectively, 
\begin{align}
S_{\rm{P}} &= \frac{1}{2}\left[ 1 + \tanh \left( \frac{\Delta P_{\rm{c}}-\Delta P_{\rm{c}}^{\rm{target}}}{\Delta P_{\rm{width}}} \right) \right], \nonumber \\
S_{\rm{B}} &= \frac{1}{2}\left[ 1 + \tanh \left( \frac{\Delta E_{\rm{r}}-\Delta E_{\rm{r}}^{\rm{target}}}{\Delta E_{\rm{width}}} \right) \right], \nonumber\\
S_{\rm{C}} &= \rm{max} \left( \frac{\rm{CAH}^{\rm{cutoff}} - {\rm{CAH}} }{\rm{CAH}^{\rm{cutoff}}} , 0 \right),
\label{eqn:score}
\end{align}
from which the total score is the geometric mean of these: $\rm{Score} = \left(S_{\rm{P}} S_{\rm{B}} S_{\rm{C}} \right) ^{\frac{1}{3}}$. For $S_{\rm{P}}$ and $S_{\rm{B}}$, a tanh profile is selected to appropriately localise critical pressures and energy barriers within a range of suitable operating conditions. Thus leads to the widths chosen here as $\Delta P_{\rm{width}} = 0.5$ and $\Delta E_{\rm{width}} = 5 \times 10^{-5}$.  Meanwhile the linear function for $S_{\rm{C}}$ aims to ensure that low-CAH structures are always favoured. 

By maximising this 6-dimensional scoring function using either the results from the wetting property survey, or a genetic algorithm, the optimal structure is obtained with a score of 0.794. The optimal parameters are ($A_{\rm{r}}$, $H_{\rm{r}}$,$L_{\rm{r}}$, $t_{\rm{r}}$, $W_{\rm{r}}$, $B/B_{\rm{ref}}$) = (0.05, $\geq$0.17, 0.17, 0.05, 0.27, 0.32). The optimum system scale of 320 nm is strikingly similar to that of springtail cuticles \cite{Nickerl2013}. Both the springtail cuticle and membrane design have been selected for pressure-resistant liquid shedding ability, whilst allowing the unimpeded movement of gasses through the surface. The membrane design proposed here may therefore reflect a natural optimum for robust gaseous diffusion. The optimum design yields the properties: $\Delta P_{\rm{c}}$ = 162 kPa, $\Delta E$ = 1.25$\times$10$^{3}$ k$_{\rm{B}}$T, CAH = 42$^{\circ}$ ($\theta_{\rm{a}} = 165^{\circ}$, $\theta_{\rm{r}} = 123^{\circ}$). This CAH is typical of currently manufactured reentrant microtextures, see for example \cite{Tuteja2007, Grigoryev2012, Zhao2012}. 

By studying the individual wetting properties, we can rationalise the optimal structural design. The optimal value of $A_{\rm{r}}$ represents the minimum pillar width tested, whose sole function is to maximise $\Delta E_{\rm{r}}$. $H_{\rm{r}}$ reflects the observation that the maximum critical pressure is achieved at $H_{\rm{r}} \geq H_{\rm{c}}$. $L_{\rm{r}} + t_{\rm{r}}$ optimises the CAH, whilst the specific value of  $L_{\rm{r}}$ maximises $\Delta E_{\rm{r}}$. Finally, the small value of $W_{\rm{r}}$ reduces the CAH, whilst retaining a high $\Delta P_{\rm{c}}$ due to the small system scale.
 
The scoring function at fixed optimal values of $A_{\rm{r}}$, $H_{\rm{r}}$, and $t_{\rm{r}}$ is shown in Fig. \ref{fig:SIM_OPT}(a)(i) as a 3D contour plot. A 2D cut through this is shown in Fig. \ref{fig:SIM_OPT}(a)(ii) at the optimal $L_{\rm{r}}$, to show that the optimal structure scale is bounded by the critical pressure criterion from above, and the minimum energy barrier criterion from below. Also shown in Fig. \ref{fig:SIM_OPT}(a)(ii) are projections of successive generations of the genetic algorithm. The optimal structure was located after 20 generations, requiring the sampling of only 0.01\% of the 7.2$\times$10$^6$ possible structures considered overall.

The four examples of manufactured doubly reentrant surfaces feature system sizes of 1-100 $\upmu$m, as smaller scales are currently challenging to manufacture \cite{Choi2017,Liu2014,Liu2018,Yun2018}. We now choose to optimise a structure whose manufacture has already been demonstrated, so we fix $B = 100$ $\upmu$m in accordance with the texture designed in a recent work by Liu and Kim \cite{Liu2014}. A leading-edge application for this is surfaces designed for digital microfluidics \cite{Freire2016,Ding2012}. Devices fail when droplets become immobilised by pinning to the surface, particularly problematic for low surface tension solvents \cite{Ding2012}, or for reactive processes where the surface tension is variable and hard to predict. 

Both of these problems are readily overcome using the doubly reentrant geometry. We demonstrate this by optimising a surface structure for use within a particularly challenging scenario - digital microfluidics using microlitre volume droplets of n-hexane ($\upgamma_{\rm{lv}}$ = 27.4 mN$\cdot$m$^{-1}$). The pressure within such a droplet (88 Pa) introduces the target pressure for use in the scoring scheme Eq. \eqref{eqn:score}: $\Delta P_{\rm{c}}^{\rm{target}}$ = 100 Pa, with the width $\Delta P_{\rm{width}} = 0.005$. Furthermore, as the CAH should be minimised, but impose the condition $\rm{CAH}^{\rm{cutoff}}$ = 50$^{\circ}$. At the imposed length scale used, the barrier score $S_{\rm{B}} \approx 1$ ($\Delta E$ is of the order of 10$\times$10$^{10}$ k$_{\rm{B}}$T), meaning that we choose to optimise the score $(S_{\rm{P}}S_{\rm{C}})^{1/2}$. 

Overall, the optimum structure, 0.508, is obtained using both the wetting property survey, and the genetic algorithm. The optimum parameters are ($H_{\rm{r}}$,$D_{\rm{r}}$, $W_{\rm{r}}$) = ($\geq$0.13, 0.22, 0.16), yielding the properties: $\Delta P_{\rm{c}}$ = 105 Pa, CAH = 25$^{\circ}$. The pillar width $A_{\rm{r}}$ and ratio of $L_{\rm{r}}$ to $t_{\rm{r}}$ become free parameters to choose. A 2D contour plot of the scoring function at constant $H_{\rm{r}} = 0.3$ is shown in Fig. \ref{fig:SIM_OPT}(b), in which projections of successive generations of the genetic algorithm are shown. Here, the algorithm converged after 14 generations, requiring 2.2\% of the entire population to be sampled.

For both the membrane distillation and digital microfluidics applications, the sensitivity of the optimised structural dimensions and properties can be assessed relative to the choice of scoring function parameters. This is achieved through re-optimising the geometries when each parameter in the scoring functions shown in Eq. \eqref{eqn:score} were varied individually by $\pm$5\%. It is found that the optimised membrane distillation geometry is insensitive to the parameter variation. This observation also applies to the optimised digital microfluidics geometry, except in the scenario where $\rm{CAH^{cutoff}}$ is reduced by 5\%. In this case, the optimal CAH is reduced by 13\%, and the optimal critical pressure is reduced by 17\%. However, this variation is due to the $\rm{CAH^{cutoff}}$ reduction (2.5$^{\circ}$) being relatively large compared to the low optimal CAH (25$^{\circ}$) for this application.

The manufactured structure reported by Liu and Kim \cite{Liu2014} is similar to the optimum geometry. However, the key difference is that the optimal geometry has a significantly shorter pillar height $H_{\rm{r}}$ than the manufactured geometry by a factor of 3.7 times. This is due to the surprisingly short critical height required, as discussed in the Critical Pressure section.

\section{\textbf{Conclusions and Outlook}}

Overall, in order to optimise the wetting properties of the reentrant and doubly reentrant surface texture for a vast variety of potential applications, we began by developing three computational strategies in order to comprehensively survey the key surface wetting properties: the contact angle hysteresis, critical pressure, and minimum energy barrier to the wetting transition. This was achieved for both wetting and non-wetting liquids (shown in SI).

In the contact angle hysteresis study, we identified four major receding mechanisms, of which only two had previously been reported, and defined the structural dimensions where each is operative. For all receding mechanisms, we were able to develop and analyse quantitative models which were robustly validated against our simulation results.

In the critical pressure study, three failure mechanisms were observed and quantified as a function of the structural parameters. However, upon comparison with the simulation data, the prevailing and widely-used critical pressure models were found to be significantly over-simplified. This lead to a particularly poor description of the liquid-vapour interface morphology, meaning that manufactured structures are many times taller (and mechanically weaker) than necessary. By developing a more sophisticated model, we were able to achieve both quantitative accuracy of the critical pressures, and success at modelling the complex interface morphologies as capillary bridges. 

In the minimum energy barrier study, we identified three failure mechanisms, quantified each barrier across the structural parameter space, and assessed which mechanism was most likely for a given geometry. Crucially, we showed how the doubly reentrant geometry is prone to condensation within the cap, but also deduced effective designs to mitigate against this.

Finally, we found that the structural features which tend to maximise the critical pressure, minimise the energy barrier, and maximise the contact angle hysteresis. As it was not possible to optimise a surface geometry with respect to each individual wetting parameter, we performed the optimisation by considering all three simultaneously. This was achieved in two ways for the optimum design of both membranes for water-purification, and surfaces for digital microfluidics. Firstly, using the comprehensive wetting property surveys, we were able to evaluate and locate the maximum of a combined scoring function. However, we then demonstrated that a genetic algorithm was able to efficiently locate the optimum design in the six-dimensional parameter space. Although the designs tested here featured a relatively small number structural degrees of freedom, going forward we highlight such optimisation techniques as being powerful tools in designing more complex structures for special wettability applications. The computational techniques developed here are highly versatile, and can be used for any mesoscopically structured surface in contact with multiple fluid phases. 

Coupled with recent, significant developments in fabrication techniques (including 3D printing, fluidization of polymer micropillars, and lithographic methods), we believe the multifaceted optimisation strategy presented here will be a powerful approach to designing real-world superomniphobic surfaces. In the future, an additional step will be to consider the mechanical reliability and scalability of manufactured designs, in which we have contributed to this discussion with our large reduction in the necessary pillar height.

\section{\textbf{Methods}}
\subsection{\textbf{Diffuse interface model}}
The simulations used to compute the CAH, critical pressure, and minimum energy barrier all employ the same diffuse interface model and system discretisation. Specific system set-ups for each wetting property are presented in SI. Within the bifluidic diffuse interface model used, the order parameter $\phi(\textbf{r})$ is chosen to represent the local composition at point \textbf{r} ($\phi = 1$ in the pure liquid phase, and $\phi = -1$ in the pure vapour phase at zero applied pressure). Based on a previous work, the free energy functional $\Psi[\phi]$ is composed of three terms \cite{Panter2017},
\begin{equation}
\Psi[\phi] = \Psi_i + \Psi_s - \Delta P V_l. \label{eqn:psi}
\end{equation}
$\Psi_i$ is the isotropic free energy term, expressed as an integral over the entire system volume $V$,
\begin{equation}
\Psi_i =\int_V \left(\frac{1}{\epsilon} \left(\frac{1}{4} \phi^4 - \frac{1}{2}\phi^2 + \frac{1}{4} \right) + \frac{\epsilon}{2} \vert \nabla \phi \vert^2 \right)dV,
\end{equation}
enforcing the equilibrium values of $\phi = \pm1$ via the double well potential, and exacting an energetic penalty for forming an interface of width $\epsilon$. This leads the the liquid-vapour surface tension $\upgamma_{\rm{lv}} = \sqrt{8/9}$.

$\Psi_i$ is the fluid-solid interaction term, expressed as an integral over the surface area,
\begin{equation}
\Psi_s = \int_S h\left( -\frac{1}{6}\phi_s^3 +  \frac{1}{2}\phi_s + \frac{1}{3} \right) dS,
\end{equation}
where $\phi_s$ is the value of $\phi$ at the surface. $h$ is the wetting parameter, and is related to the microscopic contact angle $\theta_o$ through $h = -\sqrt{2}\cos \theta_{\circ}$. The cubic wetting potential negates spurious compositional changes close to the surface by ensuring $\phi_s$ is equal to the bulk composition \cite{Connington2013}.

Within the external pressure term, the total liquid volume is calculated from
\begin{equation}
V_{\rm{l}} = \int_V \frac{\phi+1}{2} dV.
\end{equation}

The simulation system is discretised into a cubic lattice of $N_{\rm{x}} \times N_{\rm{y}} \times N_{\rm{z}}$ nodes, in which each node is either located within the solid structure, on the solid surface, or in the fluid bulk.

\subsection{\textbf{Genetic algorithm}}
We began by randomly sampling the parameter space to generate an initial population of 40 surface structures. These were ranked based on score, and the top 20 retained for breeding. Candidate pairs for breeding were selected at random, and breeding occurred if the geometric mean of their scores was greater than a random number between 0 and 1. The offspring were equally likely to inherit each attribute from either parent. Each attribute was then mutated if a random number was less than the current mutation probability $p_i$ (set initially at 0.5). For the discrete structural variables, a mutation changed the attribute randomly by between -3 and 3 lattice spacings. For the continuous variable (system scale), the change in $\log (B / B_{\rm{ref}} )$ was selected randomly from the range -0.3 and 0.3. The mutation probability was reduced each generation, such that $p_{i+1} = p_0 \times$ (standard deviation of scores in previous generation)$^{1/2}$. Any offspring bred or mutated outside the testable parameter range was mutated back into the testable parameter range. The algorithm was deemed to have converged when the mutation rate decreased to zero.

\clearpage


\part*{Supplementary information}

\section{Extended methods}

\subsection{\textbf{Contact angle hysteresis}}
To simulate the advancing and receding scenarios, we adapted the microchannel setup used by Mognetti and Yeomans \cite{Mognetti2010}. In this, a single column of surface structures patterned the base of the microchannel with microscopic contact angle $\theta_{\circ} = 60^{\circ}$. The base of each individual structure was centred within  a square of $N_{\rm{x}} \times N_{\rm{y}} = 60 \times 60$ lattice spacings, where each lattice spacing is equal to the interface width $\epsilon$. Thus, in units of  $\epsilon$ the system size $B=60$. The top of the channel was capped with a smooth surface of variable contact angle $\theta$, and was set above the top of the surface structures at a height of 40 lattice units. This height was chosen to ensure that the top surface did not influence the advancing and receding interface morphologies, which is evidenced as the interface being planar at the top surface. 

\begin{figure}[!ht]
\includegraphics[width=0.5\textwidth]{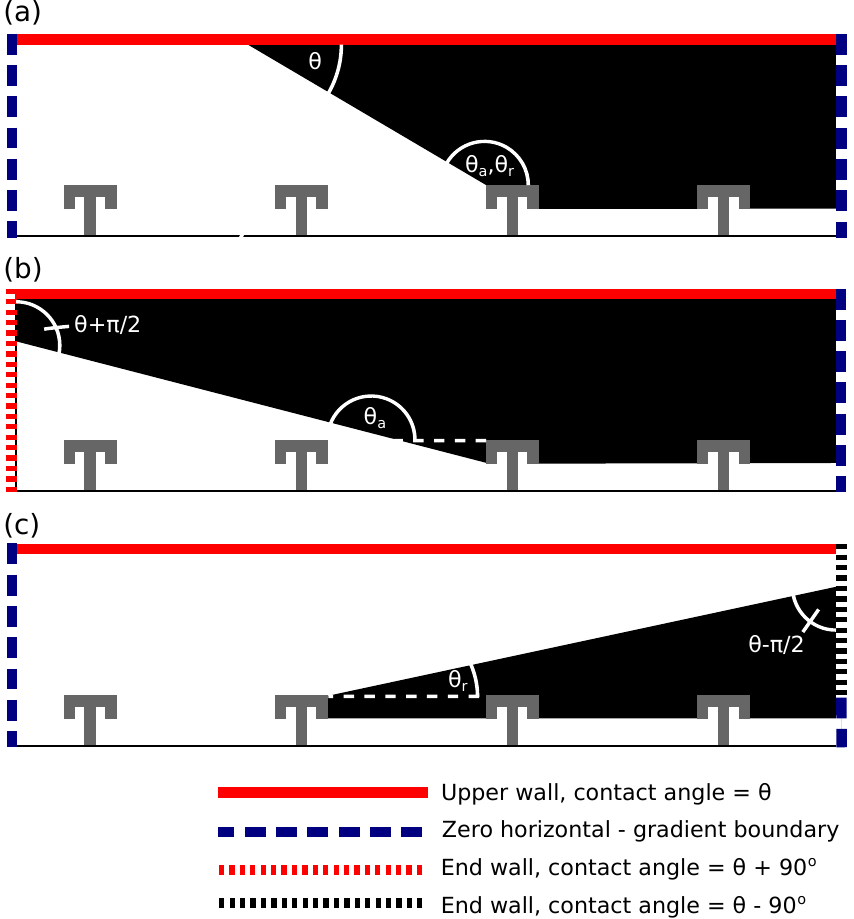}
\caption{Microchannel boundary conditions for: (a) the conventional setup, (b) the advancing case where $\theta < 20^{\circ}$, (c) the receding case where $\theta > 160^{\circ}$.}
\label{fig:1boundaries}
\end{figure}

For the majority of CAH simulations, the boundary conditions employed were as shown in Fig. \ref{fig:1boundaries}(a). Here, the gradient in $\phi$ perpendicular to the end walls was fixed at zero, enforcing bulk fluid behaviour. To determine $\theta_{\rm{a}}$, an iterative process was used in which $\theta$ was incremented and the free energy minimised via the L-BFGS algorithm \cite{Liu1989, Kusumaatmaja2015}, up to the point at which the advancing interface morphology was obtained to a precision of $0.1^{\circ}$. At convergence, the advancing contact angle $\theta_{\rm{a}} = 180^{\circ} - \theta$. An identical process was used to determine the receding interface morphology. At convergence, the receding contact angle $\theta_{\rm{r}} = 180^{\circ} - \theta$. The relationships between $\theta_{\rm{a}}$, $\theta_{\rm{r}} $, and $\theta$ are indicated in Fig. \ref{fig:1boundaries}(a).

Special modifications to the microchannel were required when advancing at $\theta < 20^{\circ}$, or receding at $\theta > 160^{\circ}$, as the diffuse interface caused a reduction in contact angle accuracy. In the former case, as shown in Fig. \ref{fig:1boundaries}(b), the advancing-end boundary was replaced by a wall with contact angle equal to $\theta + 90^{\circ}$. As before, $\theta_{\rm{a}} = 180^{\circ} - \theta$.
In the latter case, as shown in Fig. \ref{fig:1boundaries}(c), the receding-end boundary was partially replaced by a wall with contact angle equal to $\theta - 90^{\circ}$.  As before, $\theta_{\rm{r}} = 180^{\circ} - \theta$. The height at which this wall started was between the lip bottom and top, so as not to artificially affect the interface morphology, or receding contact angle. In both cases, the end walls were located sufficiently far from the contact line pinning location that the interface was planar in the end wall vicinity. 
 
\subsection{\textbf{Critical pressure}}

For each critical pressure simulation, a double-resolution system was employed in which the solid structure was centred within a simulation volume of $N_{\rm{x}} \times N_{\rm{y}} \times N_{\rm{z}} =  240 \times 240 \times (H+D+16) $ where the lattice spacing was equal to $\epsilon/2$. Thus, in units of $\epsilon$, the system size $B=120$. This was used to ensure that the contact line pinning at the critical pressure was not sensitive to the diffuse interface width. As the critical pressure interface morphologies never broke the square symmetry of the system, computational efficiency was increased by simulating one quarter of the system: $ x \in [0, N_{\rm{x}}/2], y \in [0, N_{\rm{y}}/2], z \in [0, N_{\rm{z}}]$. Mirror boundary conditions at $x = 0, N_{\rm{x}}/2$ and $y = 0, N_{\rm{y}}/2$ maintained the square symmetry. The $z$-gradient at $z = N_{\rm{z}}$ was set to zero to enforce bulk fluid behaviour. Each system was initiated in a zero-pressure suspended state, then an iterative process was used to find the critical pressure by an iterative process of pressure increase and energy minimisation. The largest $\Delta P_{\rm{r}}$ at which the suspended state remained stable was determined with a precision of $0.01 \upgamma_{\rm{lv}}/B$.

\subsection{\textbf{Minimum energy barriers}}

Here, we develop a rapid and precise transition state search method, suitable for any surface design, which can be readily implemented in pre-existing minimisation routines.

The gradient-squared method employed to survey the transition states shares similarities to that used recently for atomistic simulations \cite{Duncan2014}. The free energy minima are found by locating the stationary points of $\Psi$ at which all eigenvalues are positive. If we transform the landscape such that we minimise the gradient-squared, $\Gamma$,
\begin{equation}
\Gamma = \sum\limits_{ijk} \left( \frac{\partial \Psi}{\partial \phi_{\rm{ijk}}}	\right)^2,
\end{equation}
then all stationary points become minima of $\Gamma$, with $\Gamma = 0$. Inflection points also become minima of $\Gamma$, but have $\Gamma \neq 0$. Since the transformed landscape exhibits a large number of basins of attraction, care must be taken to initialise the system close to a transition state. This was was achieved through using the Doubly Nudged Elastic Band algorithm \cite{Trygubenko2004} once for each transition type to locate an approximate transition state. An effective method was to use these as system initialisations, and by making small changes to the surface structure, we could ensure that we never escaped the basin of attraction of the transition state (where it existed), and so reliably converged on the transition state upon minimising $\Gamma$. 

Each transition state was obtained on a single structural replica within a cubic simulation volume of $N_{\rm{x}} \times N_{\rm{y}} \times N_{\rm{z}} =  60 \times 60 \times 60 $ where the spacing was equal to $\epsilon$. Thus, the system size $B=60$. We limited the range of geometries tested as we imposed a minimum $C_{\rm{r}}\rm{(min)} = 0.1$ to ensure the diffuse interface did not dominate the wetting behaviour, thus $A_{\rm{r}} < W_{\rm{r}} - 2C_{\rm{r}}\rm{(min)}$.

\section{Extended discussion for $\theta_{\rm{\circ}} = 60^{\circ}$}
\subsection{Advancing and receding contact angles}

\begin{figure}[!ht]
\includegraphics[width=0.5\textwidth]{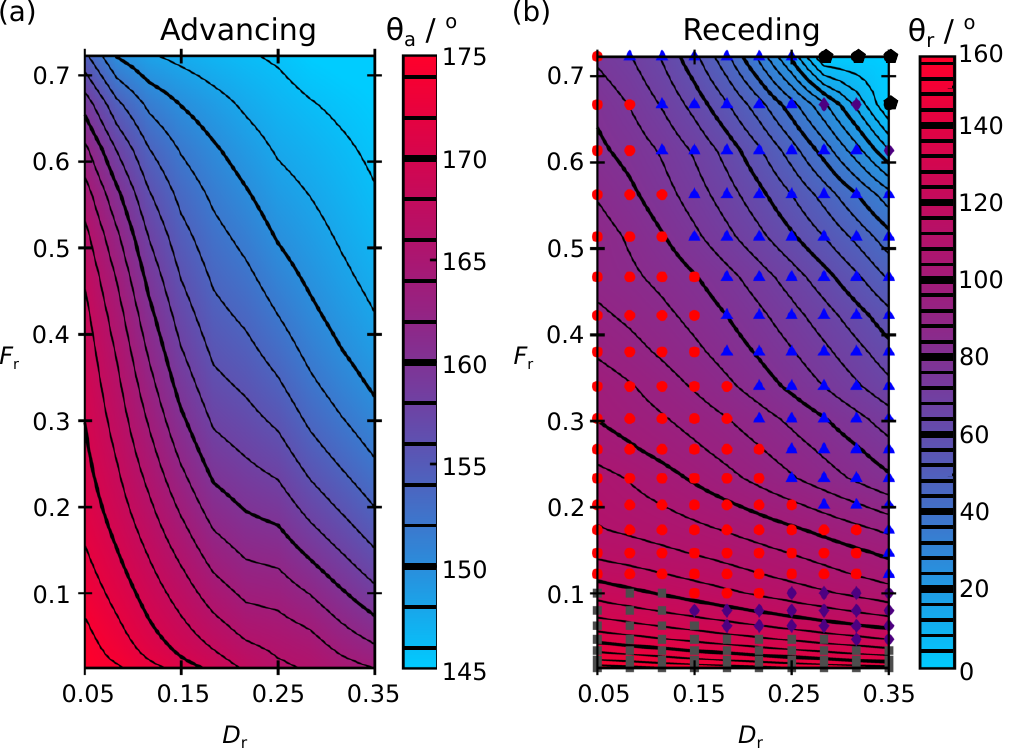}
\caption{Contour plots of the individual advancing (a) and receding (b) contact angles, as the total lip depth $D_{\rm{r}}$ and area fraction  $F_{\rm{r}}$ are varied. For the receding case, the receding mechanisms are indicated: bridge depinning (grey squares), edge depinning (red circles), lip depinning (blue triangles), non-depinning (black pentagons), and hybrid-depinning (purple diamonds).}
\label{fig:2ADVREC}
\end{figure}

In Fig. \ref{fig:2ADVREC}, we show the advancing and receding contact angles measured from simulations separately. For all practical surfaces used in superomniphobic applications, the advancing contact angle $\theta_{\rm{a}}$ exceeds $160^{\circ}$, shown in Fig. \ref{fig:2ADVREC}(a). $\theta_{\rm{a}} < 160^{\circ}$ is only observed for large lip depths and area fractions, both of which are effective at pinning the advancing interface to the advancing bottom edge of the cap. Although limited, a direct comparison with previous experimental results can be made at $F \approx 0.196$, $D_{\rm{r}} =$ 0.087 and 0.17, in which octane was found to make a microscopic contact angle with surface of $60^{\circ}$  \cite{Zhao2012}. Experimentally,  $\theta_{\rm{a}} \approx 157^{\circ}$ and $152^{\circ}$ for each $D_{\rm{r}}$ respectively, compared to $\theta_{\rm{a}} = 169^{\circ}$ and $163^{\circ}$ measured in this simulation. The 7\% difference between each value is negligible compared to the experimental precision of measuring such large contact angles. 

The receding contact angles shown in Fig. \ref{fig:2ADVREC}(b) span almost the entire possible contact angle range. We highlight the small variation in $\theta_{\rm{r}}$ with $D_{\rm{r}}$ at small area fractions $F_{\rm{r}}$, compared to the large variation in $\theta_{\rm{a}}$. It is these contrasting variations which are responsible for the optimal values of $D_{\rm{r}}$ (or $L_{\rm{r}}$) due to the CAH, presented in the simultaneous optimisation. We now compare the simulated $\theta_{\rm{r}}$ with the same experiments as discussed for $\theta_{\rm{a}}$ \cite{Zhao2012}. At $D_{\rm{r}} =$ 0.087 and 0.17, experimentally $\theta_{\rm{r}} \approx 108^{\circ}$ and $99^{\circ}$ respectively. Via simulation, we observe $\theta_{\rm{r}} = 108^{\circ}$ and $103^{\circ}$ respectively. Within the error margins ($\pm 2^{\circ}$ in experiment, 
$\pm 1^{\circ}$ in simulation), the experimental and simulation results are in very good agreement.

\subsection{Receding models}
\subsubsection{Bridge depinning}

\begin{figure*}[!ht]
\includegraphics[width=\textwidth]{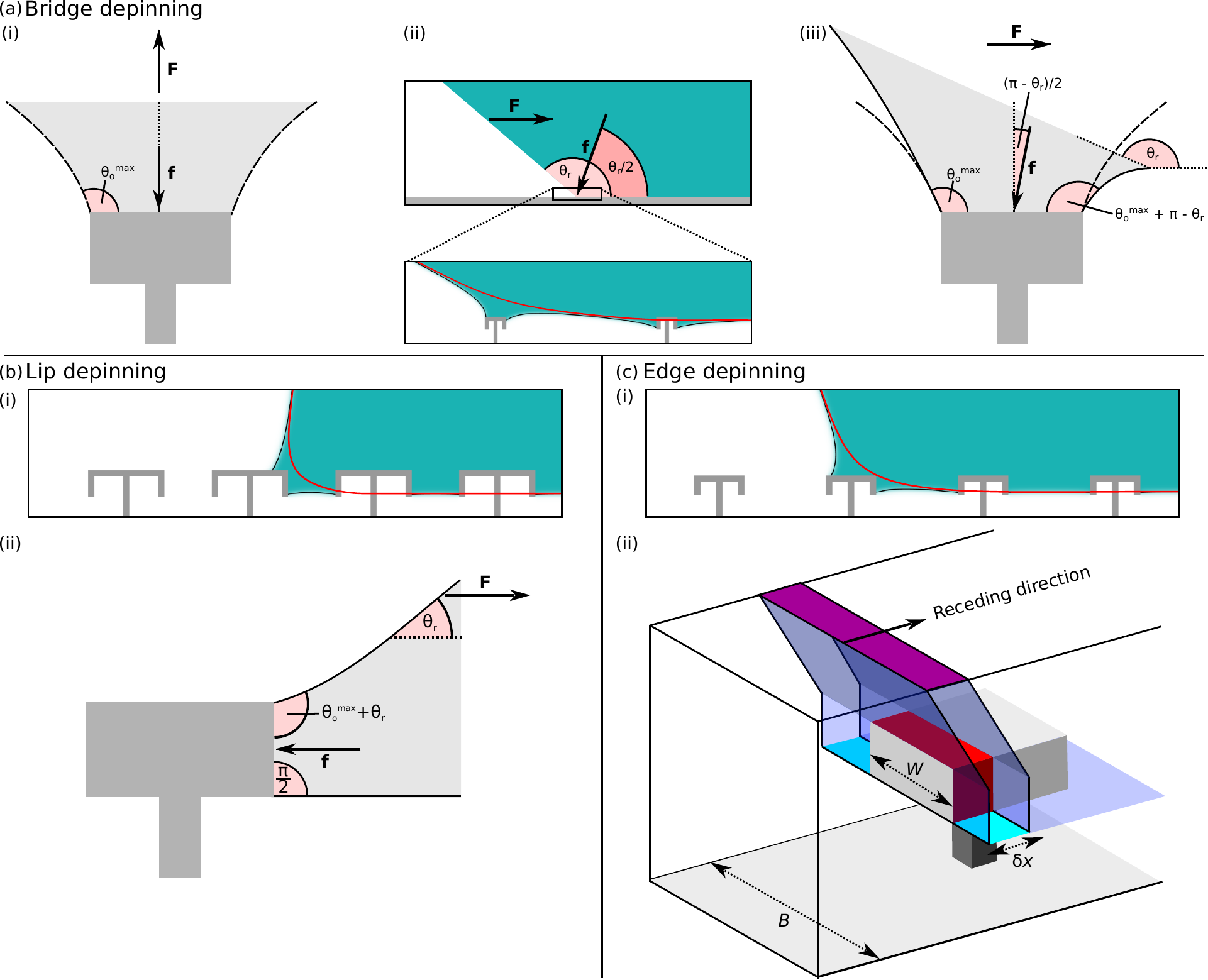}
\caption{(a) Bridge depinning model. (i) Force balance occurring at the point of depinning for a vertically strained capillary bridge. The solid reentrant geometry is shown in dark grey, the liquid in pale grey. The system is axisymmetric about the vertical dotted line. (ii) Macroscopic view of the receding interface (top), with magnification about the three-phase contact line to observe the bridge depinning mechanism (bottom). (iii) Model of a single capillary bridge being strained in the receding direction. Dashed black lines indicate the original liquid-vapour interface shape for the vertically strained bridge in (i), whilst the solid black lines indicate the strained interface shape. (b) Lip depinning model. (i) A 2D slice along the centre of the microchannel showing the lip depinning mechanism, in which the liquid (blue) is pinned to the side of the cap. (ii) Model of the lip depinning mechanism in which the liquid meniscus (pale grey) is strained parallel to the receding direction. (c) Edge depinning model. (i) A 2D slice along the centre of the microchannel showing the edge depinning mechanism. (ii) Model of the edge depinning mechanism in which the liquid vapour interface (translucent blue) is displaced by a distance $\delta x$ along the length of the microchannel. The areas highlighted in purple, red, and cyan, indicate the change in liquid contact area with the top surface, cap, and vapour phase respectively. }
\label{fig:3depin}
\end{figure*}

In order to model the bridge depinning mechanism, we begin by considering the maximum pinning force a capillary bridge can achieve when being strained parallel to it's axis, illustrated in Fig. \ref{fig:3depin}(a)(i). In this, we show a 2D slice through the centre of an axisymmetric system. The axis of symmetry is denoted by the vertical dotted line. Assuming the contact angle at the three phase contact line is uniform throughout, at the point of depinning the maximum pinning force $\mathbf{f}$ is expressed as
\begin{equation}
|\mathbf{f}| = \upgamma_{\rm{lv}} m \cos \theta_{\rm{\circ}}^{\rm{max}}, \label{eqn:pinforce}
\end{equation}
where $m$ is the contact line length, and $\theta_{\rm{\circ}}^{\rm{max}}$ is the corrected contact angle.  $\theta_{\rm{\circ}}^{\rm{max}}$ is defined as $\theta_{\rm{\circ}}^{\rm{max}} = \max[\theta_{\rm{\circ}}, \uppi/2]$.

In receding systems, the capillary bridge is strained along the receding direction, shown in Figs. \ref{fig:3depin}(a)(ii,iii). Following a previous derivation \cite{Butt2015}, but augmented with our treatment of $\theta_{\rm{\circ}}^{\rm{max}}$, two effects arise. Firstly the direction of the pinning force $\mathbf{f}$ balances the capillary forces arising from the front and receding edge of the cap, and so forms an angle which bisects the receding contact angle \cite{Butt2013}, shown in Fig. \ref{fig:3depin}(a)(ii). Secondly, the contact angle around the contact line varies between a maximum value at the innermost edge (right hand side of Fig. \ref{fig:3depin}(a)(iii)), to a minimum value at the outermost edge (left hand side of Fig. \ref{fig:3depin}(a)(iii)). At the innermost edge, the receding angle increases the local contact angle to  $\theta_{\rm{\circ}}^{\rm{max}} + \uppi - \theta_{\rm{r}}$. At the outermost edge, the contact angle cannot be reduced from $\theta_{\rm{\circ}}^{\rm{max}}$, otherwise the contact line would depin. The pinning force $\mathbf{f}$ is assumed to depend on the average contact angle $\bar{\theta}$, approximated as $\bar{\theta} =  \theta_{\rm{\circ}}^{\rm{max}} + \left(\uppi - \theta_{\rm{r}}\right) / 2.$

At equilibrium, the horizontal component of the microscopic pinning force $\mathbf{f}_{\rm{h}}$ must balance the macroscopic receding force $\mathbf{F}$, leading to
\begin{align}
|\mathbf{f}_{\rm{h}}| &= \upgamma_{\rm{lv}} m \sin \bar{\theta} \sin \left(  \frac{\uppi}{2} -  \frac{\theta_{\rm{r}}}{2}   \right), \nonumber \\
&=  \upgamma_{\rm{lv}} m \cos \left(  \theta_{\rm{\circ}}^{\rm{max}} -  \frac{\theta_{\rm{r}}}{2}   \right) \cos \frac{\theta_{\rm{r}}}{2}.
\end{align}
On the macroscale, the force $\mathbf{F}$ required to displace a contact line of length $B$ is expressed via the Young-Dupr\'e equation for the work of adhesion: $|\mathbf{F}| = \upgamma_{\rm{lv}}B\left( 1 + \cos \theta_{\rm{r}} \right)$. Thus, by equating $|\mathbf{f}_{\rm{h}}| = |\mathbf{F}|$ and manipulating the equality, we arrive at a general expression for the receding contact angle for liquids of all wettabilities:
\begin{equation}
\tan\frac{\theta_{\rm{r}}}{2} = \frac{2B}{m\sin\theta_{\rm{\circ}}^{\rm{max}}} - \frac{1}{\tan\theta_{\rm{\circ}}^{\rm{max}}}. \label{eqn:CAH_bridge}
\end{equation}
In the case of wetting liquids, for which $\theta_{\rm{\circ}}^{\rm{max}} = \frac{\uppi}{2}$, this simplifies to
\begin{equation}
\tan\frac{\theta_{\rm{r}}}{2} = \frac{2}{4W_{\rm{r}}\alpha},
\end{equation}
where we have substituted the reduced perimeter $m/B$ for that appropriate for square caps: $m/B = 4W_{\rm{r}}\alpha$.

\subsubsection{Lip depinning}
In the lip depinning scheme, the geometry of the capillary-bridge depinning mechanism is markedly altered, shown in Fig. \ref{fig:3depin}(b). At zero applied pressure, the liquid-vapour interface pinned to the bottom of the caps is planar, making a contact angle of $\uppi/2$ with the cap side. Thus, the microscopic pinning force $\mathbf{f}$ acts anti-parallel to the macroscopic receding force $\mathbf{F}$, as shown in Fig. \ref{fig:3depin}(b)(ii). The effect of $\mathbf{F}$ is therefore to change the contact angle around the contact line, varying from a maximum value ($\theta_{\circ}^{\rm{max}}+\theta_{\rm{r}}$) at the top of the cap, to a minimum value ($\uppi/2$) at the bottom of the cap. We make the assumption that $\mathbf{f}$ depends on the average contact angle  $\bar{\theta} =  \frac{\theta_{\rm{\circ}}^{\rm{max}}}{2} +  \frac{\theta_{\rm{r}}}{2} + \frac{\uppi}{2}$, leading to
\begin{equation}
|\mathbf{f}| = \upgamma_{\rm{lv}}m \sin \left( \frac{\theta_{\rm{\circ}}^{\rm{max}}}{2} +  \frac{\theta_{\rm{r}}}{2} + \frac{\uppi}{2} \right).
\end{equation}
For wetting liquids with $\theta_{\rm{\circ}}^{\rm{max}} = \frac{\uppi}{2}$, this simplifies to 
\begin{equation}
|\mathbf{f}| = \upgamma_{\rm{lv}}m \cos  \frac{\theta_{\rm{r}}}{2}.
\end{equation}
Finally, we equate $|\mathbf{f}|$ and $|\mathbf{F}|$ to yield an expression for the receding contact angle due to lip depinning,
\begin{equation}
\cos \frac{\theta_{\rm{r}}}{2} = \frac{2(W_{\rm{r}}+D_{\rm{r}})\alpha}{2},
\end{equation}
where we have substituted the reduced perimeter $m/B$ for that appropriate for rectangular  caps sides: $m/B = 2(W_{\rm{r}}+D_{\rm{r}})\alpha$. We note that for non-wetting surfaces ($\theta_{\circ} > \uppi/2$), as the liquid-vapour interface does not pin to the cap underside in the suspended state, the lip depinning mechanism cannot occur.

\subsubsection{Edge depinning}
The final model we consider is that of edge depinning for wetting liquids, illustrated in Fig. \ref{fig:3depin}(c). We approximate the mechanism as both the liquid-vapour interface and three-phase contact line sliding without changing shape. Thus, we are able to evaluate the energy change for this process over a small displacement $\delta x$ of the contact line along the microchannel. The energy change is expressed as the sum of contributions arising from sliding across the top surface of the microchannel, sliding across the surface of the cap, and reducing the liquid-vapour interfacial area, such that
\begin{align}
\delta E = & \upgamma_{\rm{sv}}^{\rm{top}} B \delta x - \upgamma_{\rm{sl}}^{\rm{top}} B \delta x \nonumber \\
&+ \upgamma_{\rm{sv}}^{\rm{cap}} \left( W + 2D \right) \delta x - \upgamma_{\rm{sl}}^{\rm{cap}} \left( W + 2D \right) \delta x \nonumber \\
&-\upgamma_{\rm{lv}} \left( B-W \right) \delta x.
\end{align}  
This is readily simplified using the Young equation to yield an expression for the force required to exact the displacement of the contact line,
\begin{equation}
\frac{\delta E}{\delta x} = \upgamma_{\rm{lv}} \left[ B\cos \theta_{\rm{top}} + (W+2D)\cos \theta_{\circ} - (B-W) \right]. 
\end{equation}
At the point of receding, this force is equal to zero. By realising that $\theta_{\rm{r}} = \uppi - \theta_{\rm{top}}$, this yields
\begin{equation}
\cos \theta_{\rm{r}} = \left( W_{\rm{r}} + 2D_{\rm{r}} \right) \cos \theta_{\circ} + W_{\rm{r}} -1. \label{eqn:CAH_edge}
\end{equation}

\subsection{Critical pressure models}
\subsubsection{Critical pressure derivation}

\begin{figure}[!ht]
\includegraphics[width=0.5\textwidth]{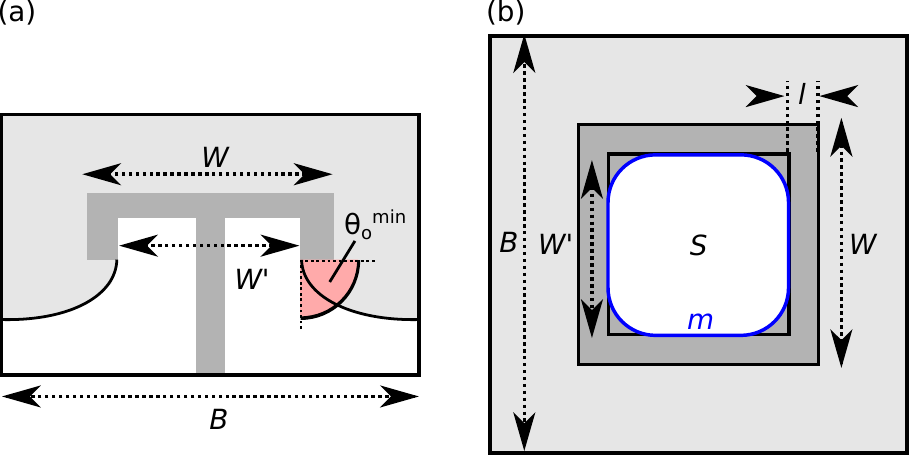}
\caption{(a) Side view of a 3D system at the critical pressure. In this, the liquid (pale grey) sags beneath the cap (dark grey),and the contact line is pinned with contact angle $\theta_{\circ}^{\rm{min}}$. (b) Underside view of the same system, showing the position of the contact line perimeter of length \textit{m} (blue line), and projected area occupied by the vapour phase, S. \textit{W}' indicates the pinned width of the contact line (in general different to \textit{W}).}
\label{fig:4critp}
\end{figure}
To model the critical pressure, we improve on the derivation presented by Tuteja \textit{et al.} \cite{Tuteja2007, Tuteja2008} to capture the actual contact line morphology. We begin by considering a typical system at the critical pressure, of size \textit{B}, cap width \textit{W}, illustrated in Fig. \ref{fig:4critp}. At the critical pressure, the pinning force of the contact line balances the force due to the pressure, such that
\begin{equation}
\left( \Delta P_{\rm{c}} \upgamma_{\rm{lv}} / B \right)\left( B^2 - S\right) =  \upgamma_{\rm{lv}} m \sin \theta_{\circ}^{\rm{min}}, \label{eqn:pcrit2}
\end{equation} 
where $\Delta P_{\rm{c}}$ is the reduced critical pressure (normalised with respect to $\upgamma_{\rm{lv}} / B$). $\left( B^2 - S\right)$ is the projected area occupied by the liquid phase, shown in Fig. \ref{fig:4critp}(b). As in the contact angle hysteresis study, \textit{m} is the contact line perimeter, and again we take care to correctly capture the true pinning force by using the corrected contact angle, where $\theta_{\circ}^{\rm{min}} = \min[\theta_{\circ}, \uppi/2]$ if the contact line is pinned to the outer cap edge, or $\theta_{\circ}^{\rm{min}} = \uppi/2$ if the contact line is pinned to the inner cap edge. We note that the former case is appropriate for wetting on (singly) reentrant geometries, or doubly reentrant geometries where $\theta_{\circ} > \uppi/2$. The latter case is appropriate for doubly reentrant geometries where $\theta_{\circ} < \uppi/2$. In all cases, we make the assumption that the average contact angle is equal to $\theta_{\circ}^{\rm{min}}$. 

To model the contact line perimeter, we assume that the pinned width of the contact line is $W'$, which is different from the cap width $W$ by a distance $a$. For pinning on the inner lip (for wetting liquids on doubly reentrant structures as shown in Fig. \ref{fig:4critp}), we expect $a$ to be approximately twice the lip width $l_{\rm{r}}$. For pinning on the outer lip (for reentrant geometries, or non-wetting liquids on doubly reentrant geometries), we expect $a \approx 0$. Overall, $m = 4\upalpha W'$, where the shape parameter $\upalpha$ is used to smoothly vary the possible perimeter shapes from circular ($\upalpha=\uppi/4$) to square ($\upalpha=1$). To model the projected vapour-occupied area $S$, we again use the corrected width $W'$ and shape parameter $\upalpha$ to model the contact line shape, yielding $S = \upalpha W'^2$. The critical pressure model is obtained by manipulating Eq. \eqref{eqn:pcrit2},
\begin{align}
\Delta P_{\rm{c}} &= B \sin \theta_{\circ}^{\rm{min}} \frac{m}{B^2-S}, \nonumber \\
&= \sin \theta_{\circ}^{\rm{min}} \frac{4\upalpha W' B}{B^2 - \upalpha W'^2}, \nonumber \\
&= \sin \theta_{\circ}^{\rm{min}} \frac{4\upalpha}{\frac{1}{W_{\rm{r}}'} - \upalpha W_{\rm{r}}'}, \label{eqn:pcrit_model}
\end{align}
where $W_{\rm{r}}'$ is the reduced corrected width $W'/B$.

\subsubsection{Critical height derivation}
\begin{figure*}[!ht]
\includegraphics[width=\textwidth]{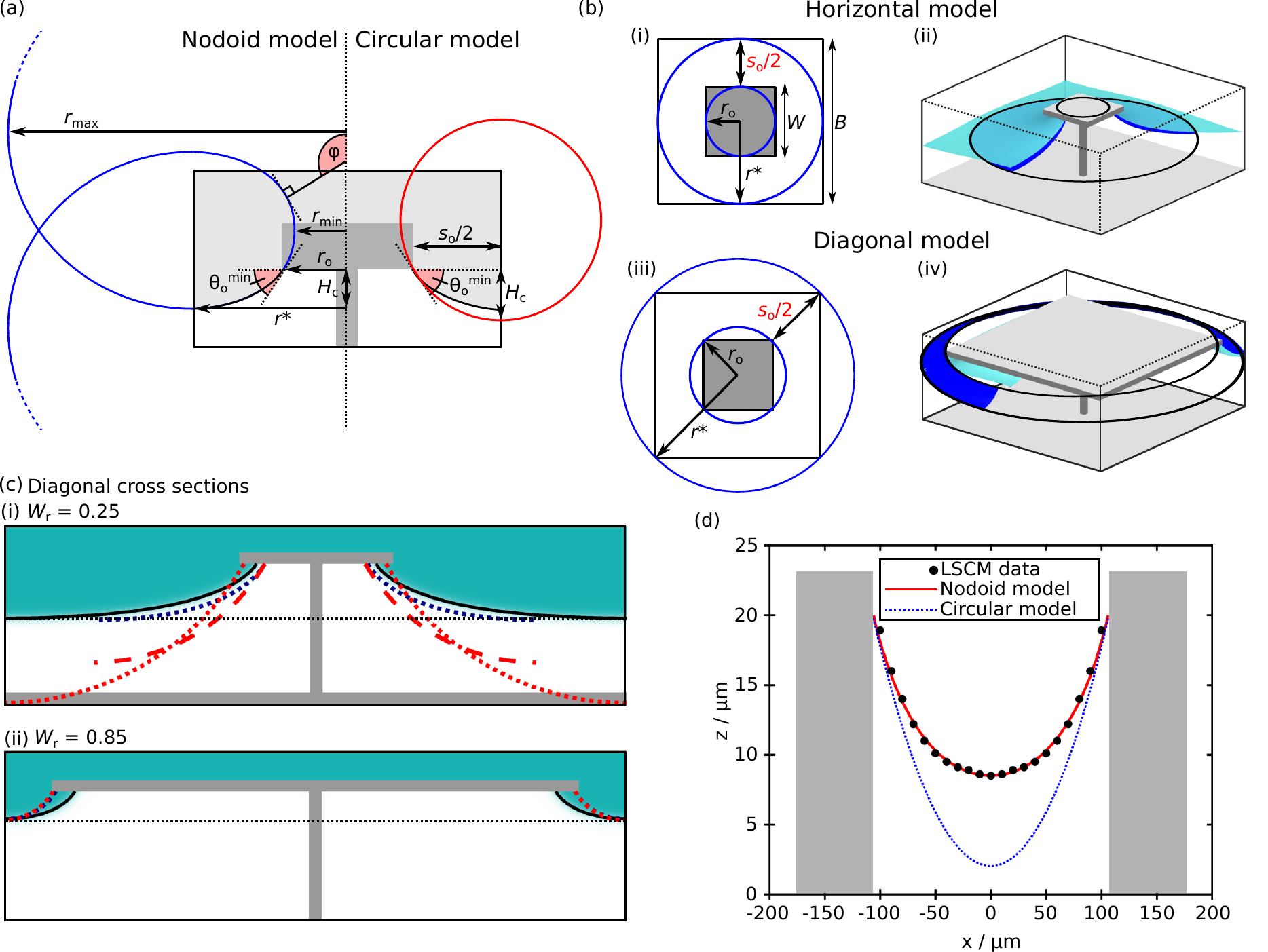}
\caption{(a) Models used to approximate the true liquid-vapour interface morphology (black line). A reentrant geometry is shown for visual clarity, but the models are also valid for doubly reentrant geometries. (Left) The nodoid model for approximating the vapour phase as a capillary bridge (blue line); (right) the circular arc model (red line). All distances and angles used in the model constructions are labelled. (b) In a square system, two choices of distance measurement exist, either horizontal (i,ii) or diagonal (iii,iv). (i,iii) vertical projections of the system, showing inner and outer radii for the nodoid model (blue), and cap separation for the circular model. (ii,iv) Example comparisons with simulations for the horizontal and diagonal models at $W_{\rm{r}}=0.25$ and $W_{\rm{r}}=0.85$ respectively. The simulated liquid-vapour interface is shown in light blue, the nodoid model shown in dark blue. Both interfaces are cut horizontally and diagonally to be able to compare the model and simulation results. Black lines indicate the inner and outer perimeters of the nodoid interface. (c) Diagonal cross sections from the 3D horizontal (i) and vertical (ii) models, the sagging height of the liquid-vapour interface obtained from simulation is indicated by a black dotted line. Also shown are the liquid vapour interfaces from: simulation (black, solid), nodoid model (blue, dotted), diagonal circular model (red, dotted), and horizontal circular model (red, dashed, only shown in (i)). In (ii), the diagonal nodoid and circular models overlap, showing how both agree closely with the simulated critical height. (d) Comparison of the diagonal nodoid (red, solid line) and circular arc models (blue, dashed line) with an experimental laser scanning confocal microscopy (LSCM) diagonal profile (black data) of a liquid-vapour interface on a square array of square pillars (grey). Experimental data extracted from \cite{Papadopoulos2013}. Note the stretched vertical scale for visual clarity.}
\label{fig:5crith}
\end{figure*}

By far the most commonly used approximation for liquid-vapour interfaces in 3D systems is the circular arc model, illustrated in Fig. \ref{fig:5crith}(a). For visual clarity we show a reentrant geometry, although the following models and derivations are general for both reentrant and doubly reentrant geometries at all contact angles. In this, a circular arc spans the separation between two adjacent pillars. At the critical pressure for cap failure the arc makes a contact angle of $\theta_{\circ}^{\rm{min}}$ with each cap underside. To choose which pillars should be spanned by a circular arc, there exist two choices, shown in Fig. \ref{fig:5crith}(b), either the arc spans horizontally separated pillars, or diagonally separated pillars. These are labelled as the horizontal and diagonal models respectively. The distance spanned by the circular arc, $s_{\circ}$, is therefore equal to $B-W'$ in the horizontal model, or $\sqrt{2}(B-W')$ in the diagonal model. Generally expressed, the radius of curvature, $R$ is
\begin{equation}
R=\frac{s_{\circ}}{2 \sin \theta_{\circ}^{\rm{min}}},
\end{equation}  
and the sagging height $H_{\rm{c}}$ is
\begin{equation}
H_{\rm{c}}=\frac{s_{\circ}}{2}\frac {1-\cos \theta_{\circ}^{\rm{min}}}{\sin \theta_{\circ}^{\rm{min}}}. \label{eqn:HCRIT_2D}
\end{equation} 
We emphasise that the sagging height at the onset of cap failure is equivalent to the critical height, as a geometry with a height $H = H_{\rm{c}}$ would fail simultaneously via cap failure and base failure.

In Fig. \ref{fig:5crith}(c), the circular arc model is tested for accuracy at representing the simulated interface shape. In Fig. \ref{fig:5crith}(c)(i) ($W_{\rm{r}}=0.25$), the commonly used diagonal circular arc model (dotted red line) grossly overestimates the sagging height at the critical pressure by several times. The horizontal circular arc model also overestimates the sagging height (dashed red line). It is only at the largest cap widths that the circular arc is able to accurately estimate $H_{\rm{c}}$, as shown for the diagonal model in Fig. \ref{fig:5crith}(c)(ii) ($W_{\rm{r}}=0.85$). The reason for this 2D-like behaviour at large $W_{\rm{r}}$, is that the principal radius of curvature approximated by the circular arc ($R_1$) is significantly smaller than the second principal radius of curvature of the interface ($R_2$). This arises because $s_{\circ} << W_{\rm{r}}$. Thus, the Laplace pressure $\Delta P \propto 1/R_1 + 1/R_2$ and hence the interface shape, becomes well-approximated by the single radius of curvature $R_1$.

However, surfaces with large $W_{\rm{r}}$ suffer from a severely limited capacity to produce surfaces of low contact angle hysteresis. For practically useful surfaces where the circular arc model is highly inaccurate, we consider an alternative capillary-bridge model. In this, we recognise that the liquid-vapour interface under the cap forms a capillary-bridge-like structure, in which the vapour has a negative pressure relative to the liquid. We therefore look to model the interface as the simplest 3D surface of constant mean curvature. Out of the family of Delaunay surfaces, the nodoid exhibits a negative mean curvature. This is illustrated in Fig. \ref{fig:5crith}(a), in which the blue line shows a portion of interfacial profile of the axisymmetric, periodic surface. In the parametrisation shown, the vertical height $z$ of a point on the surface is defined by the local radius $r$ and angle relative to the surface-normal $\varphi$. In the following derivation, we only present the nodoid characteristics pertinent to ascertaining the critical height, for a comprehensive treatment see for example \cite{Krotov1999}. The surface is fully characterised by specifying the innermost radius, $r_{\rm{min}}$, outermost radius, $r_{\rm{max}}$, and a single point on the surface. This point can be determined by realising that, at the critical pressure, the interface makes an angle of $\uppi - \theta_{\circ}^{\rm{min}}$ in the vapour phase with the underside of the cap at the pinning location. Thus, we are able to specify the point ($z_{\circ}$, $r_{\circ}$, $\varphi_{\circ}$) = ($H$, $W'/2$, $\theta_{\circ}^{\rm{min}}$) for the horizontal nodoid model, or ($H$, $\sqrt{2}W'/2$, $\theta_{\circ}^{\rm{min}}$) for the diagonal nodoid model, shown in Fig. \ref{fig:5crith}(b). We are now able to deduce the appropriate $r_{\rm{min}}$ and $r_{\rm{max}}$ for each pillar geometry. Firstly, on the portion of the nodoid representing the liquid-vapour interface, there exists a point at radius $r^{*} = \sqrt{r_{\rm{min}}r_{\rm{max}}}$ where $dr/dz = 0$. To respect the periodic boundary conditions of the liquid-vapour interface, $r^{*} = B/2$ or $r^{*} = \sqrt{2}B/2$ in the horizontal and diagonal models respectively. Secondly, by rearranging the relationship
\begin{align}
\sin \varphi &= \frac{r^2 - r_{\rm{min}}r_{\rm{max}}}{r(r_{\rm{max}} - r_{\rm{min}})}, \nonumber \\
&= \frac{r^2 - r^{*2}}{r(r^{*2}/r_{\rm{min}} - r_{\rm{min}})},
\end{align}
we are able to find $r_{\rm{min}}$ as
\begin{equation}
r_{\rm{min}} = \frac{1}{2r\sin\varphi}\left[ r^{*2}-r \pm \sqrt{ \left( r-r^{*2}\right)^2 + \left(rr^{*}\sin\varphi\right)^2} \right],
\end{equation}
by taking the negative result and substituting for one of the defined points $(r,\varphi) = (r_{\circ},\varphi_{\circ})$.

Finally, we are able to obtain the nodoid approximation to the interfacial profile through the relationship
\begin{align}
&z(r) =  r_{\rm{min}} F(k,\psi) - r_{\rm{max}} E(k, \psi),  \\
&\rm{where} \quad \mathit{k}^2 = 1-\frac{\mathit{r}_{\rm{min}}^2}{\mathit{r}_{\rm{max}}^2}, \\
&\rm{and} \quad  \sin^2 \psi = \frac{\mathit{r}_{\rm{max}}^2 - \mathit{r}^2}{\mathit{r}_{\rm{max}}^2-\mathit{r}_{\rm{min}}^2},
\end{align}
which is plotted in Fig. \ref{fig:5crith}(c) (blue dashed line). $F(k,\psi)$ and $E(k,\psi)$ are the elliptic integrals of the first and second kinds respectively. Overall, $H_{\rm{c}}$ is obtained through recognising that
\begin{equation}
H_{\rm{c}} = z(r_{\circ}) - z(r^{*}). \label{eqn:HCRIT_3D}
\end{equation}  

Experimental validation of the nodoid model is shown in Fig. \ref{fig:5crith}(d). Here, experimental data (black points) are extracted from laser scanning confocal microscopy (LSCM) experiments performed in \cite{Papadopoulos2013}, which was able to resolve a diagonal cross-section of the liquid-vapour interfacial profile of a sessile water droplet on a square array of square pillars. For this system, the pillar width $W$ = 50 $\mu$m, height $H$ = 23 $\mu$m, system size $B$ = 200 $\mu$m, and the contact angle is determined to be $109^{\circ}$. As the nodoid and circular arc models only require a knowledge of the contact angle, pillar separation and contact line diameter, both models are equally applicable for simple pillars, reentrant, and doubly reentrant geometries. We compare diagonal variants of both models with the diagonal experimental profile in Fig. \ref{fig:5crith}(d). The nodoid model (red, solid line) is observed to closely follow the experimental data, whereas the circular arc model (blue, dashed line) overestimates the sagging height of the interface by 56\%.

\subsection{Transition mechanisms}
\subsubsection{Individual mechanisms}
\begin{figure*}[!ht]
\includegraphics[width=\textwidth]{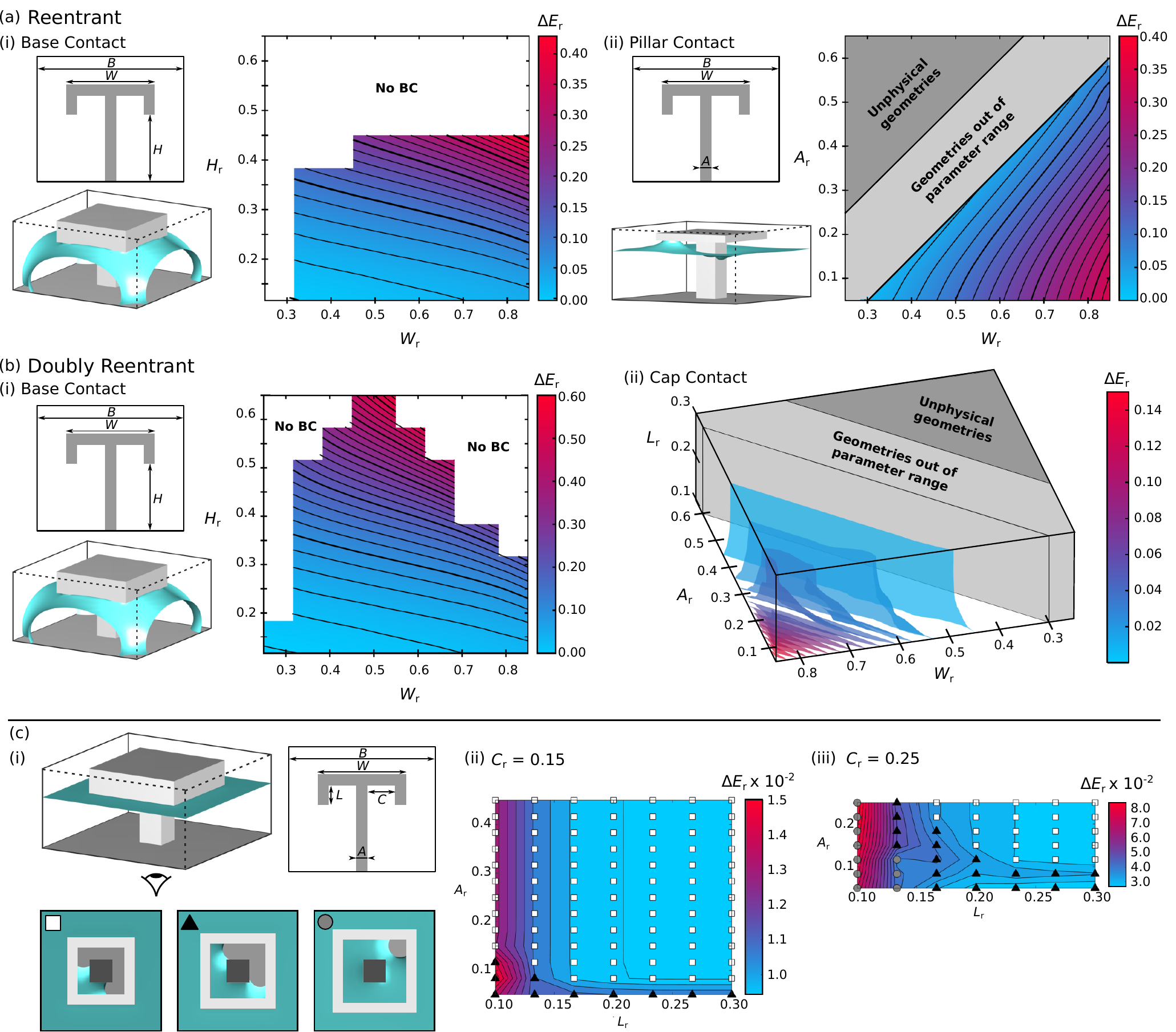}
\caption{(a) 3D illustration and contour plot of the two reentrant transition mechanisms: (i) Base Contact, (ii) Pillar Contact. Impossible geometries with pillar widths $A_{\rm{r}}$ wider than the cap width $W_{\rm{r}}$ are shaded in dark grey. Geometries approaching this limit and requiring infeasibly large computational domains are shaded in light grey. (b) 3D illustration and contour plot of the two doubly reentrant transition mechanisms: (i) Base Contact, (ii) Cap Contact. (c) Extended description of the cap contact transition mechanism. (i) 3D illustration and under-cap views of the three transition variants (each labelled with a square, triangle, or circle). (ii, iii) Two slices through the 3D contour plot at constant cap width values $C_{\rm{r}}=0.15$ and $C_{\rm{r}}=0.25$ respectively. Each datum is labelled with a symbol corresponding to the transition variant shown in (i). }
\label{fig:6ts}
\end{figure*}

To conclude the extended discussion for the reentrant and doubly reentrant geometries at $\theta_{\circ} = 60^{\circ}$, we present the energetic barriers for each separate wetting mechanism in Fig. \ref{fig:6ts}. 

For the reentrant geometry, the two possible mechanisms are the Base Contact and Pillar Contact modes, shown in Fig. \ref{fig:6ts}(a). For Base Contact, at large pillar heights $H_{\rm{r}}$, the three-phase contact line is no longer able to remain pinned to the base of the cap, and is unstable with respect to depinning and sliding inwards. This places an upper height limit on the existence range of Base Contact, shown in Fig. \ref{fig:6ts}(a)(i). This upper height limit decreases with decreasing cap width $W_{\rm{r}}$ in a manner similar to the suppressed critical height at low $W_{\rm{r}}$ in the critical pressure discussion. In this, the small, negative radius of curvature of the liquid-vapour interface around the cap enforces an even smaller, positive radius of curvature outwards from the cap. The effect of this is to increase the contact angle the liquid-vapour interface makes with the cap underside, meaning that sliding occurs at smaller pillar heights $H_{\rm{r}}$ than for structures with larger cap widths $W_{\rm{r}}$.

For Pillar Contact on the reentrant geometry, the energetic barrier does not depend on the pillar height $H_{\rm{r}}$, and is therefore the only possible transition mechanism at large pillar heights where Base Contact is inoperative. However, two parameters which do influence the Pillar Contact barrier are the cap width $W_{\rm{r}}$ and pillar width $A_{\rm{r}}$. The energetic barrier is increased when the area of the liquid-vapour interface at the transition state is increased. This is observed to occur by maximising $W_{\rm{r}}$ and minimising $A_{\rm{r}}$, as shown in Fig. \ref{fig:6ts}(a)(ii).

For the doubly reentrant geometry, the two transition mechanisms, Base Contact and Cap Contact,  are shown in Fig. \ref{fig:6ts}(b). For cap widths $W_{\rm{r}} < 0.7$, the Base Contact mechanism, shown in Fig. \ref{fig:6ts}(b)(i), is operative over a greater range of pillar heights $H_{\rm{r}}$ than for the reentrant geometry. This is due to the inner cap lip being an effective pinning site, so preventing the liquid-vapour interface from sliding inwards. However, at cap widths $W_{\rm{r}} > 0.7$, it becomes energetically favourable for the liquid-vapour interface to depin from the cap corners, leading to liquid filling the underside of the cap. In these scenarios, Base Contact becomes unstable with respect to the cap-filling, Cap Contact mechanisms.

For the doubly reentrant geometry, the Cap Contact mechanism exists over the entire parameter range tested, shown in Fig. \ref{fig:6ts}(b)(ii). Because Cap Contact is a heterogeneous condensation mechanism, in which nucleation proceeds from one cap corner, we expect Cap Contact to be operative for all physical doubly reentrant geometries. As with pillar contact on the reentrant geometry, the energetic barrier is increased by increasing the area of the liquid-vapour interface at the transition state. This is shown in  Fig. \ref{fig:6ts}(b)(ii) to be achieved by maximising the cap width $W_{\rm{r}}$ and minimising the pillar width $A_{\rm{r}}$.

However, three mechanistic variants of Cap Contact are observed, shown in Fig. \ref{fig:6ts}(c)(i), and their existence ranges are shown in Figs. \ref{fig:6ts}(c)(ii,iii). For ease of comparing the transition mechanisms, throughout we have expressed the Cap Contact energy barrier as a function of the cap width $W_{\rm{r}}$. However, the more pertinent parameter for the internal condensing mode is the inner cap width $C_{\rm{r}}$. We show contour plots for the Cap Contact energy barrier at constant $C_{\rm{r}} = 0.15$ and $C_{\rm{r}} = 0.25$ in Figs. \ref{fig:6ts}(c)(i) and (ii) respectively. When the internal cavity width $C_{\rm{r}}$ is small, at large lip depths $L_{\rm{r}}$ and large pillar widths $A_{\rm{r}}$, liquid is most readily able to condense within the cavity as the energetic penalty for forming a liquid-vapour interface is offset early in the transition by forming a large energetically favourable liquid-solid contact area. Thus, under these conditions, the critical nucleus is relative small (labelled with the white square). As the cavity size $C_{\rm{r}}$ increases, the lip depth $L_{\rm{r}}$ decreases, or the pillar widths $A_{\rm{r}}$ decreases, the energetic offset for forming the liquid-vapour interface is reduced in magnitude. Thus, the critical nuclei occur later in the transition pathway, with larger energy barriers to overcome. These higher-energy critical nuclei are labelled with black triangles, or grey circles in the most extreme cases.

\subsubsection{Combined mechanisms}
\begin{figure*}[!ht]
\includegraphics[width=\textwidth]{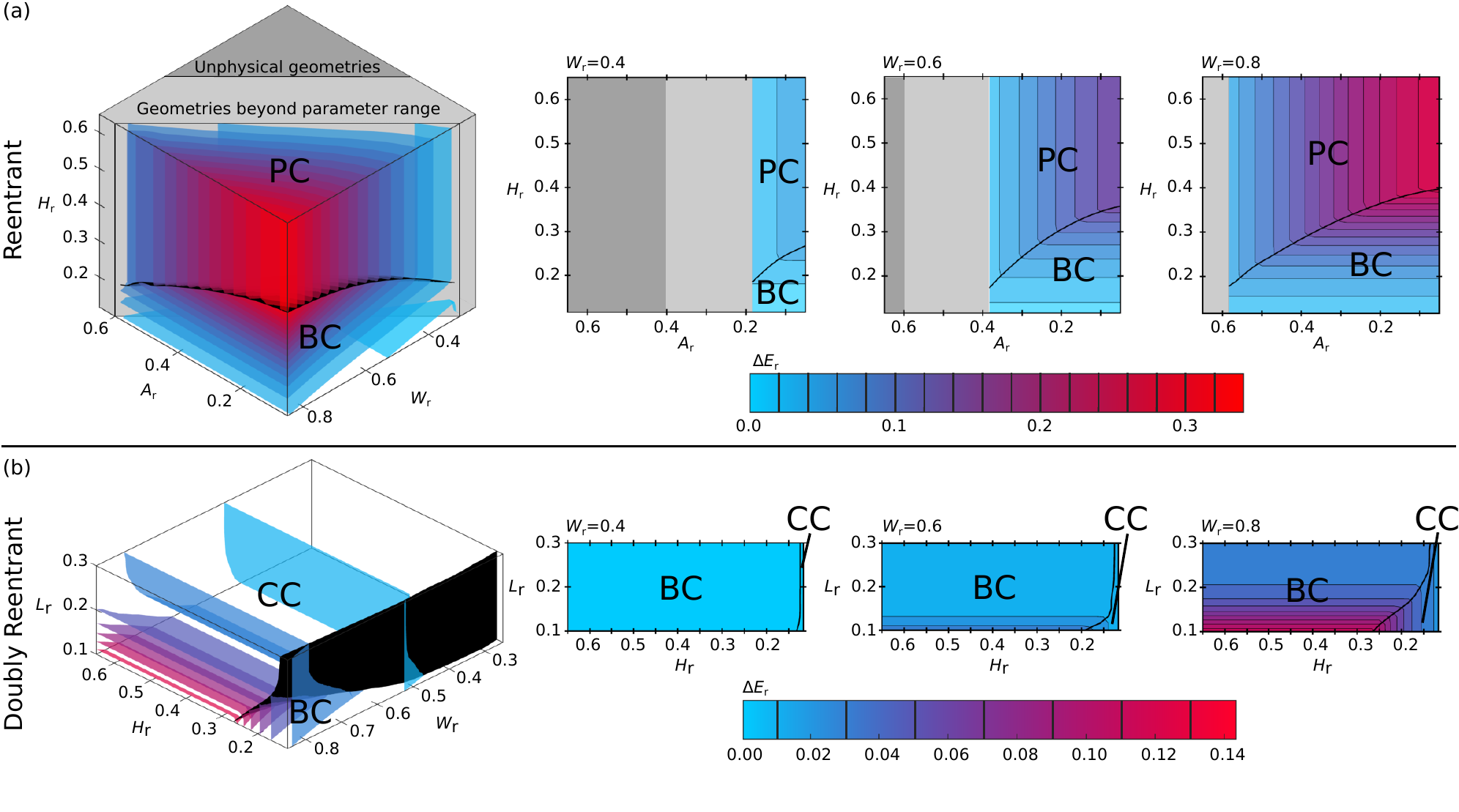}
\caption{(a) 3D contour plot of the lowest energy reentrant geometry transition mechanisms: Base Contact, BC; and Pillar Contact, PC. Three slices through the 3D contour plot are also shown for $W_{\rm{r}}$ = 0.4, 0.6, and 0.8. Impossible geometries with pillar widths $A_{\rm{r}}$ wider than the cap width $W_{\rm{r}}$ are shaded in dark grey. Geometries approaching this limit and requiring infeasibly large computational domains are shaded in light grey. (b) 3D contour plot of the lowest energy doubly reentrant geometry transition mechanisms: Base Contact, BC; and Cap Contact, CC. Three slices through the 3D contour plot are also shown for $W_{\rm{r}}$ = 0.4, 0.6, and 0.8. }
\label{fig:7tsslice}
\end{figure*}

In Fig. \ref{fig:7tsslice}(a,b), the barriers of the lowest energy transition mechanisms are shown for the reentrant and doubly reentrant geometries respectively. For visual clarity, three additional panels are included for each, which represent cuts through the 3D contour plots at constant values of $W_{\rm{r}}$ = 0.4, 0.6, and 0.8. For both geometries these panels highlight the  increasing barrier with increasing cap width $W_{\rm{r}}$ (all other parameters fixed), as well as the competition between the transition mechanisms.

\section{Results and discussions for $\theta_{\rm{\circ}} = 110^{\circ}$}
\subsection{Contact angle hysteresis}

\begin{figure*}[!ht]
\centering
\includegraphics[width=\textwidth]{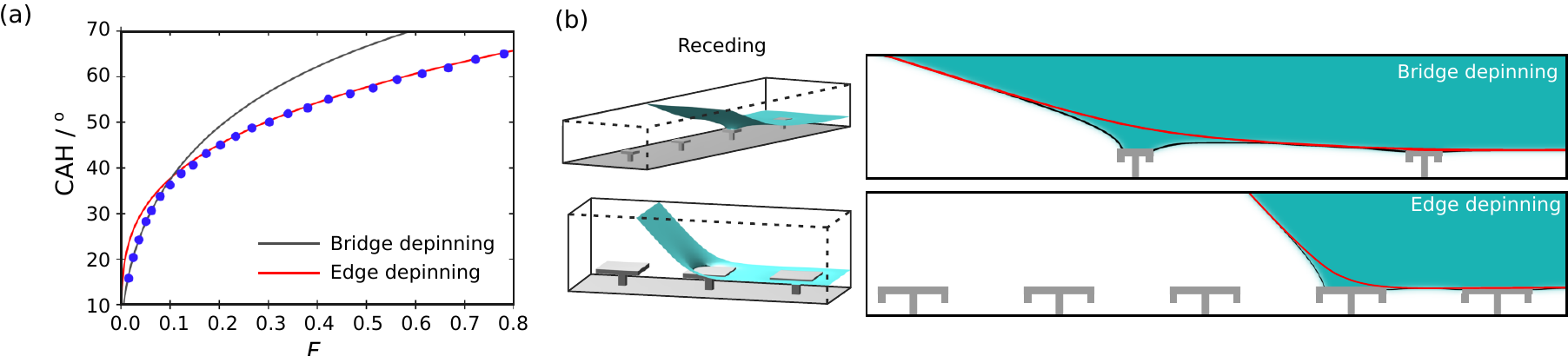}
\caption{(a) Contact angle hysteresis measured via simulation (blue circles), with fitted bridge depinning (grey line) and edge depinning (red line) models. (b) Visualisations of systems at the point of receding via the bridge and edge depinning mechanisms.}
\label{fig:8CAH110}
\end{figure*}

We now consider the wetting properties of non-wetting surfaces. Throughout, the microscopic contact angle is fixed at $\theta_{\circ} = 110^{\circ}$.

We begin by surveying the CAH, shown in Fig. \ref{fig:8CAH110}. Because the liquid-vapour interface is pinned to the top of the cap, the CAH is independent of the underlying cap structure. Thus, the reentrant and doubly reentrant geometries show identical behaviour, as do simple posts with no cap structure at all. 

The pinning at the top of the cap leads to a uniform advancing contact angle of $180^{\circ}$ for all cap widths $W_r$, matching experimental observations \cite{Schellenberger2016}. Two receding mechanisms are observed, bridge and edge depinning, shown in Fig. \ref{fig:8CAH110}(b). The receding contact angle for the bridge depinning mechanism is described by Eq. \eqref{eqn:CAH_bridge}, where the corrected contact angle $\theta_{\circ}^{\rm{max}} = \theta_{\circ}$ for  $\theta_{\circ} > \uppi/2$. Meanwhile, the receding contact angle for the edge-depinning mechanism is described by Eq. \eqref{eqn:CAH_edge}, in which the term $2D_{\rm{r}}$ is neglected as the liquid does not wet the cap sides. In this limit, Eq. \eqref{eqn:CAH_edge} reduces to the conventional receding model \cite{Choi2009}. The bridge and edge depinning mechanisms smoothly interpolate on the non-wetting geometries. We find that the error associated with both models compared to the simulation data is minimised if points with $F<0.1$ are associated with the bridge depinning model, and points with $F>0.1$ are associated with the edge depinning model. $F=0.1$ therefore corresponds to the crossover between analytic models. Overall, for the bridge model, we find $\upalpha = 0.65$ yields a maximum difference in $\theta_{\rm{r}}$ of $1^{\circ}$.  It is interesting to note that here, $\upalpha < \uppi/4$, meaning that the contact line is shorter than a circle of width $W$. Such small values of $\upalpha$ have been previously reported \cite{Butt2015}, and suggests that not all portions of the contact line contribute equally to the pinning force, as was our assumption in Eq. \eqref{eqn:pinforce}. However the excellent agreement between theory and simulation allow us to conclude that the proposed capillary bridge model provides a suitable approximation for the actual receding contact angle. The edge-depinning model also shows excellent agreement with the simulation data, in which the maximum difference in $\theta_{\rm{r}}$ between theory and simulation is also $1^{\circ}$.

\subsection{Critical pressures}
\begin{figure}[!ht]
\centering
\includegraphics[width=0.5\textwidth]{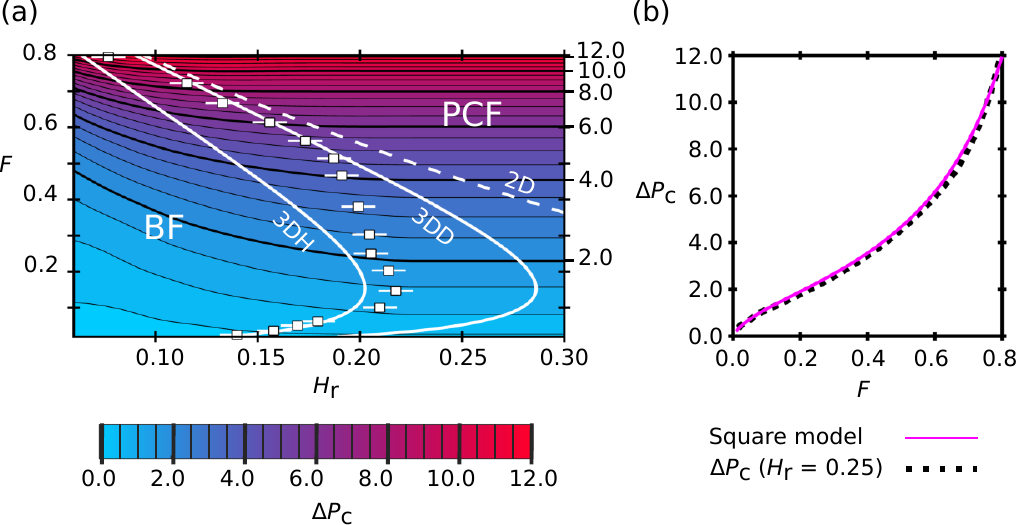}
\caption{(a) Critical pressure variation with area fraction $F_{\rm{r}}$ and pillar height $H_{\rm{r}}$. Data points indicate the critical height, and therefore the boundary between Base Failure BF, and pinned Cap Failure PCF. Error bars represent the interface width, and hence the uncertainty in the critical height. Approximations to the critical heights are indicated for circular 2D model (white, dashed line), and both horizontal and diagonal  nodoid variants (white, solid line). (b) Comparison of the cap failure model against the simulation data for $H_{\rm{r}}=0.25$.}
\label{fig:9PCRIT110}
\end{figure}

The critical pressure behaviour observed in Fig. \ref{fig:9PCRIT110} is identical for both reentrant and doubly reentrant geometries, as the contact line is always pinned to the outer cap edge. The behaviour is similar to that observed for the doubly reentrant geometry at $\theta_{\circ} = 60^{\circ}$ (pinned to the inner cap edge), and would be identical in the limit that the lip width $l \rightarrow 0$. This is because in all three cases, the corrected contact angle in Eq. \eqref{eqn:pcrit_model} is $\theta_{\circ}^{\rm{min}} = 90^{\circ}$. As found previously, the 2D circular arc approximation in Eq. \eqref{eqn:HCRIT_2D} grossly overestimates the critical height for all but the largest cap widths. The 3D capillary bridge models are again shown to be accurate at low area fractions $F_{\rm{r}}$ (horizontal model), and high $F_{\rm{r}}$ (diagonal model), and correctly capture the qualitative behaviour at intermediate values.

In Fig. \ref{fig:9PCRIT110}(b), we show the comparison between the critical pressure model for cap failure, expressed in Eq. \eqref{eqn:pcrit_model}, with the simulation data. As the contact line closely follows the cap edge, excellent agreement between the model and simulation is achieved when $\upalpha = 1$ (square contact line), yielding a fitted $a = 0.02$. As anticipated, because the contact line is pinned to the outer cap edge, $a$ is insignificantly different from zero, relative to the interface width = 0.01.

\subsection{Minimum energy barriers}

\begin{figure*}[!ht]
\centering
\includegraphics[width=\textwidth]{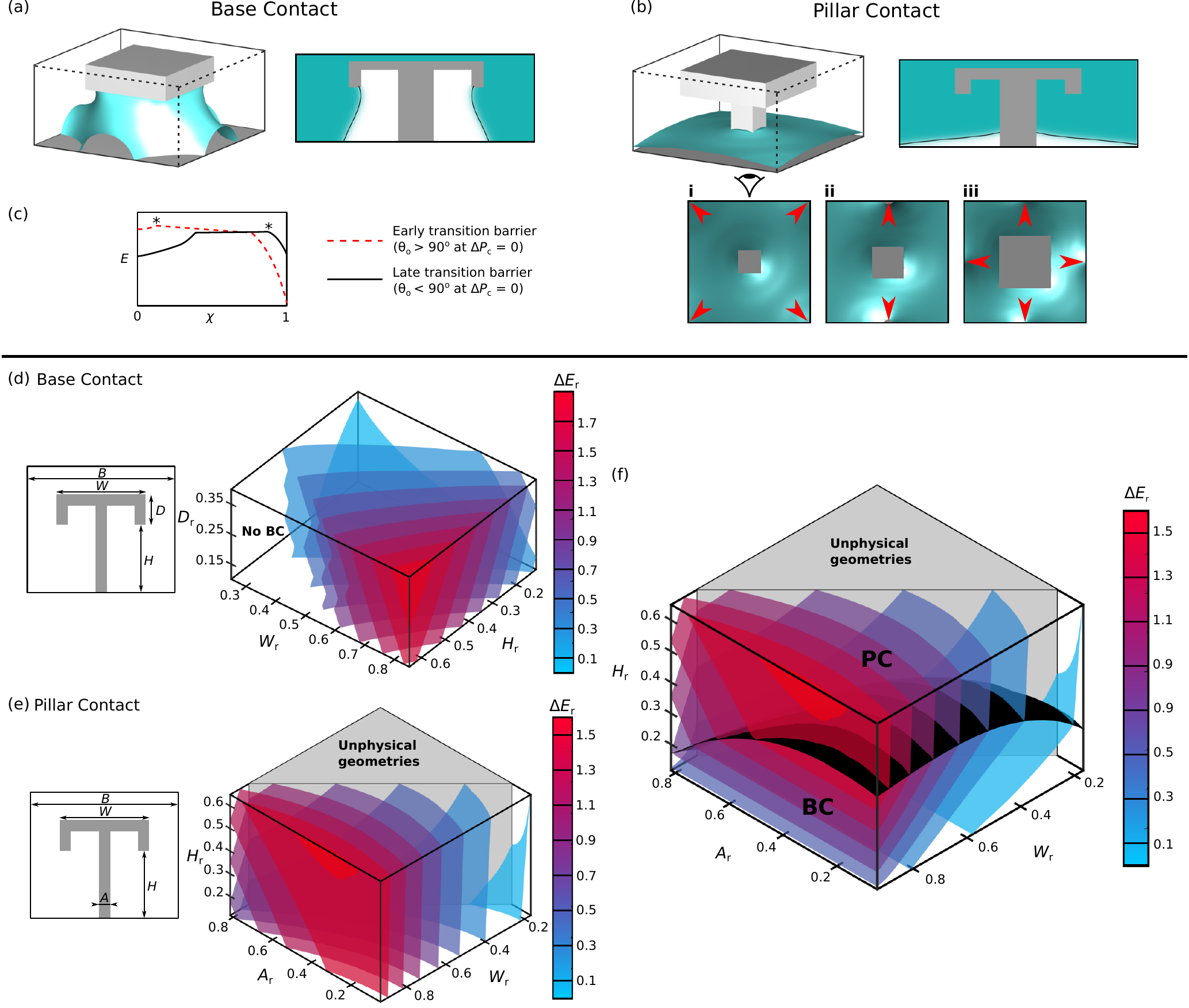}
\caption{(a) 3D visualisation and 2D diagonal cross sections of a representative Base Contact example. (b) 3D visualisation and 2D diagonal cross section a representative Pillar Contact example, in which the under-cap views shown the three symmetric variants. Red arrows indicate the positions where the liquid-vapour interface first contacts the base of the system. (c) Illustration of the two forms of Pillar Contact energetic profiles $E$ as a function of the reaction coordinate $\chi$, with barriers labelled with a *. $\chi=0$ in the suspended state and 1 in the collapsed state. (d-f) 3D contour plots showing the minimum energy barrier for Base Contact (d) and Pillar Contact (e), and the minimum energy transition path (f) as a function of the height $H_{\rm{r}}$, cap width $W_{\rm{r}}$, pillar width  $A_{\rm{r}}$, and total lip depth $D_{\rm{r}}$. Markings on the colour bars indicate the energies of each constant-barrier surface.}
\label{fig:10TS110}
\end{figure*}

For $\theta_{\circ} = 110^{\circ}$, Base Contact and Pillar Contact mechanisms are observed, illustrated in Fig. \ref{fig:10TS110}(a,b). Throughout, there is negligible difference between the wetting mechanisms for reentrant and doubly reentrant geometries. Both the Base Contact and Pillar contact mechanisms are similar to the $\theta_{\circ} = 60^{\circ}$ cases. However for Pillar Contact, as illustrated in Fig. \ref{fig:10TS110}(c), the energy barrier (*) occurs much later in the reaction pathway for $\theta_{\circ} = 110^{\circ}$ (black line) than for $\theta_{\circ} = 60^{\circ}$ (red dashed line). Despite the complex interface morphologies, a simple relationship is derived to predict which Pillar Contact variant is operative for a given set of structural parameters. If the total energy of the system is lowered by the liquid-vapour interface sliding down the pillar, the minimum energy barrier occurs early in the transition pathway. This occurs prior to sliding when the liquid contacts the cap, as observed at $\theta_{\circ} = 60^{\circ}$. Otherwise, the minimum energy barrier will occur late in the transition, after the sliding process, and immediately before the interface contacts the base of the system, as observed for $\theta_{\circ} =110^{\circ}$. Such a mechanism was originally proposed to give rise to the barrier to the Cassie-Baxter state on superhydrophobic surfaces \cite{Patankar2004}. More generally, we may consider the energy change of an interface sliding down the pillar at constant interface shape. If
\begin{equation}
\Delta P_{\rm{r}} > \frac{4\cos \theta_{\circ}}{A_{\rm{r}} - \frac{1}{A_{\rm{r}}} },
\label{eqn:PCslide}
\end{equation}
sliding is energetically favourable, and the energy barrier will occur early in the transition mechanism. As we restrict ourselves here to the study of transitions at zero pressure, this explains why the early barrier is observed for $\theta_{\circ} < 90^{\circ}$, whilst the late transition is observed for $\theta_{\circ} > 90^{\circ}$.

The Pillar Contact mechanism is further observed to have three variants, illustrated in Fig. \ref{fig:10TS110}(b). The variant is selected based on the pillar width $A_{\rm{r}}$, such that for type (i), $A_{\rm{r}} \leq 0.12$, for type (iii), $A_{\rm{r}} \geq 0.22$, and type (ii) occurs at intermediate values.

We now consider the minimum energy barrier as a function of the structural parameters. For Base Contact, shown in Fig. \ref{fig:10TS110}(d), the liquid-vapour interface morphology depends on the pillar height  $H_{\rm{r}}$ and cap width  $W_{\rm{r}}$. As the interfacial area is maximised at large $H_{\rm{r}}$ and $W_{\rm{r}}$, the largest minimum energy barriers occur under these conditions. However, upon increasing the height, there exist heights above which the contact line can no longer remain pinned to the outer cap edge, meaning that Base Contact is prevented. For non-wetting liquids, the suspended free energy minimum sees the interface pinned to the top of the cap (at zero applied pressure), whereas the liquid must wet the entire sides of the cap to overcome the minimum energy barrier. Thus, in contrast to $\theta_{\circ} = 60^{\circ}$, at $\theta_{\circ} = 110^{\circ}$, the minimum energy barrier is increased by increasing the total lip depth $D_{\rm{r}}$.

In the Pillar Contact mechanism, shown in Fig. \ref{fig:10TS110}(e), the liquid-vapour interface morphology at the minimum energy barrier is affected only by the pillar width $A_{\rm{r}}$. However, during the transition, almost the entire solid area of the reentrant or doubly reentrant pillar becomes wetted, yielding a concomitant energetic penalty. Therefore, the minimum energy barrier depends on all structural parameters, but in Fig. \ref{fig:10TS110}(e) we choose to fix $D_{\rm{r}} = 0.15$. Although the doubly reentrant cap structure has a larger surface area than the reentrant cap, for the parameter ranges considered here this only marginally impacts the minimum energy barrier.

Finally, over the range of structural parameter values, we evaluate the minimum energy transition mechanism, shown in Fig. \ref{fig:10TS110}(f). In this, we fix $D_{\rm{r}} = 0.15$. At the lowest pillar heights, the Base Contact mechanism is operative, but changes to Pillar Contact once the energetic penalty for forming the large liquid-vapour interface in Base Contact becomes too large. Unlike in the  $\theta_{\circ} = 60^{\circ}$ case, here the barrier can be increased indefinitely by increasing the total solid area of the structure which is wetted during the transition. This can be achieved effectively through extending $H_{\rm{r}}$ or $D_{\rm{r}}$.

\bibliography{Bibliography}

\section*{\textbf{Acknowledgements}}
We would like to thank Ciro Semprebon, Iris Liu, and Erte Xi for useful discussions, Carl M. Jones for updating the energy minimisation software, and P\&G and EPSRC (EP/P007139/1) for funding.

\end{document}